\documentclass[preprintnumbers,prd,showpacs,floatfix,superscriptaddress,nofootinbib, letterpaper]{revtex4-1}
\pdfoutput=1

\usepackage{longtable}
\usepackage{bm}
\usepackage{relsize}
\usepackage{amsfonts}
\usepackage{amsmath}
\usepackage{amssymb,epsf}
\usepackage{latexsym}
\usepackage{graphicx,epsfig}
\usepackage{amssymb}
\usepackage{float}
\usepackage{subfigure}
\usepackage{epstopdf}
\usepackage[colorlinks=true,citecolor=blue,linkcolor=blue,urlcolor=black]{hyperref}
\usepackage{dcolumn}
\usepackage{psfrag}
\usepackage{wrapfig}
\usepackage{makeidx}
\usepackage{epsf}
\usepackage{color}
\usepackage{multirow}
\usepackage{mathtools}

\begin{document}

\title{Junction conditions in gravity theories with extra scalar degrees of freedom}

\author{João Luís Rosa}
\email{joaoluis92@gmail.com}
\affiliation{University of Gda\'{n}sk, Jana Ba\.{z}y\'{n}skiego 8, 80-309 Gda\'{n}sk, Poland}
\affiliation{Institute of Physics, University of Tartu, W. Ostwaldi 1, 50411 Tartu, Estonia}

\date{\today}

\begin{abstract} 
In this work we present a general method to obtain the junction conditions of modified theories of gravity whose action can be written in the form $f\left(X_1,...,X_n\right)$, where $X_1$ to $X_n$ are any combination of scalar dependencies, e.g. the Ricci scalar $R$, the trace of the stress-energy tensor $T$, the Gauss-Bonnet invariant $\mathcal G$, among others. We discriminate the junction conditions into three sub-groups: the immediate conditions, arising from the imposition of regularity of the relevant quantities in the distribution formalism; the differential conditions, arising from the differential terms in the field equations; and the coupling conditions, arising from the interaction between different scalars $X_i$. Writing the modified field equations in terms of a linear combination of the different contributions of the scalars $X_i$ allows one to analyze the direct and differential junction conditions independently for each of these scalars, whereas the coupling junction conditions can be analyzed separately afterwards. We show that the coupling junction conditions induced on a scalar $X_i$ due to a coupling with a scalar $X_j$ are of the same form as the differential junction conditions of the scalar $X_j$, but applied to the analogous quantity in the framework of the scalar $X_i$. We provide a complete analysis of three different types of spacetime matching, namely smooth matching, matching with a thin-shell, and matching with double gravitational layers, and we also describe under which conditions the full sets of junction conditions might be simplified. Our results are applicable to several well-known theories of gravity e.g. $f\left(R\right)$, $f\left(T\right)$, and $f\left(\mathcal G\right)$, and can be straightforwardly extrapolated to other theories with scalar dependencies and more complicated scenarios where several dependencies are present simultaneously.
\end{abstract}

\maketitle

\section{Introduction}\label{sec:intro}

The junction conditions \cite{darmois,lichnerowicz,Israel:1966rt,papa,taub,Rosa:2023tbh} are a set of mathematical requirements that two spacetime solutions of a given theory of gravity, described by different metric tensors and valid in different sub-regions of the full spacetime manifold, must satisfy at the boundary between these two sub-regions in order to guarantee that the union of the two spacetimes is itself a valid solution of the field equations of the theory of gravity used as framework. 

In General Relativity (GR), these conditions require that the induced metric on both sides of the hypersurface that separates the two sub-manifolds coincide, and that the extrinsic curvature on both sides of that hypersurface either coincide or are associated with a thin-shell of matter \cite{Israel:1966rt}. These conditions were proven useful in accounting for a wide variety of astrophysical scenarios associated particularly with compact objects and spherically symmetric solutions \cite{schwarzschildstar,oppenheimersnyder,rosafluid,rosafluid2,Tamm:2023wvn,senovilla0,lanczos1,lanczos2,Martinez:1996ni,Lemos:2017mci,Lemos:2017aol,Lemos:2015ama,Lemos:2016pyc,brito,Mena:2017kpo}. However, the junction conditions vary depending on the theory of gravity chosen as framework, which has propelled numerous studies to derive these conditions in the framework of extended theories of gravity, e.g. $f\left(R\right)$ gravity \cite{senovilla1,Vignolo:2018eco,Reina:2015gxa,Deruelle:2007pt,Olmo:2020fri},  $f\left(R,T\right)$ gravity \cite{rosafrt,rosafrt2}, hybrid metric-Palatini gravity \cite{rosahmp}, scalar-tensor theories of gravity \cite{suffern, Barrabes:1997kk,Padilla:2012ze}, teleparallel theories of gravity \cite{delaCruz-Dombriz:2014zaa}, Einstein-Cartan theories of gravity \cite{Arkuszewski:1975fz,Luz:2019frs}, metric-affine gravity \cite{amacias}, bi-scalar Gauge gravity \cite{Casado-Turrion:2023omz}, among others. A particular application of interest of these conditions is the study of non-exotic wormhole solutions in modified theories of gravity \cite{rosaworm1,rosaworm2,rosaworm3,Rosa:2023olc,Rosa:2023guo}. 

Despite the extensive literature regarding junction conditions and their applications, there are still several potentially interesting theories of gravity for which the general sets of junction conditions have not been derived, e.g. theories of gravity depending in scalar quantities like $R_{\mu\nu}T^{\mu\nu}$, $R_{\mu\nu}R^{\mu\nu}$, $T_{\mu\nu}T^{\mu\nu}$, among others \cite{Odintsov:2013iba,Katirci:2013okf,Cognola:2012jz,Carroll:2004de}. Similarly to more popular theories like $f\left(R\right)$ or $f\left(T\right)$ gravity, the fact that these theories depend on an additional scalar quantity implies that the methods necessary to obtain the junction conditions for these theories follow a similar procedure, and thus one of the objectives of this work is to provide a unified approach on how to obtain such conditions in any theory of gravity featuring additional scalar degrees of freedom in comparison with GR through an arbitrary function.

In recent years, a large number of papers have been accepted for publication in internationally acclaimed journals, despite featuring an incorrect analysis of junction conditions and applications of the thin-shell formalism. The incorrectness of such manuscripts arises from several different origins: some manuscripts recur to the Darmois junction conditions in theories where they are not applicable \cite{Bhatti:2023wxy,Yousaf:2023jxt,Yousaf:2022duo,Yousaf:2023qtf,Yousaf:2023jpi,Farwa:2022zho,Bamba:2023evj,Bhatti:2022qye,Yousaf:2020juc,Maurya:2021mqx,Yousaf:2021yse,Yousaf:2018jkb,Ilyas:2017bmb,Bhatti:2021ryy,Bamba:2023wok,Yousaf:2023zjo}; others recur to the Israel equation to compute the stress-energy tensor of the thin-shell in theories where it is not applicable \cite{Bhatti:2020cjz,Yousaf:2021vsb,Bhatti:2022mhv,Bhatti:2022xxd,Bhatti:2021dvp,Yousaf:2021avg,Yousaf:2020srr,Yousaf:2019zcb}; some works use a set of junction conditions for a given theory that is incomplete \cite{Yousaf:2022lel,Yousaf:2021mlq,Yousaf:2020nza,Yousaf:2020dsx,anotheryousaf}; and others start from the correct complete set of junction conditions but fail to correctly derive the junction conditions for particular forms of the action \cite{Bhatti:2023cgw,Bhatti:2023mez} or evaluate only a subset of those conditions explicitly, thus failing to provide a complete proof that the solutions considered can be matched \cite{Yousaf:2022uwy,Yousaf:2023npv}. In some occasions, the manuscripts even refer to the works where the correct set of junction conditions for the theory under study has been derived, but proceed to perform the matching without recurring to these conditions \cite{Bhatti:2021oyn}. These incorrect approaches consequently render the resulting content of limited physical and mathematical relevance, e.g., certain manuscripts go as far as to mention the necessity of a thin-shell to perform the matching, even in theories where thin-shells are prohibited \cite{Yousaf:2021xex,Yousaf:2022bwc}.

The situation mentioned above serves as a hint to a lack of fundamental understanding and an indication that much caution should be taken on the topic of junction conditions and thin-shells. The main aim of this work is thus to provide a comprehensible tool on how to obtain the adequate junction conditions for several types of modified theories of gravity featuring extra scalar degrees of freedom in comparison to GR, and hopefully to suppress this gap in the literature and contribute to the improvement of the mathematical soundness and, consequently, the physical relevance and impact of future works in the field.

This work is organized as follows. In Sec. \ref{sec:theory}, we introduce a general theory of gravity for which the action depends explicitly on several scalar quantities and obtain the field equations for such a theory; in Sec. \ref{sec:dist} we introduce the formalism of distribution functions and write the necessary geometrical and matter quantities to obtain the junction conditions in this formalism; in Sec. \ref{sec:juncgen} we obtain the junction conditions for smooth matching and matching with a thin-shell, for an arbitrary form of the action that describes the theory; in Sec. \ref{sec:juncpart} we analyze a few particular cases that could lead to exceptions in the complete set of junction conditions; in Sec. \ref{sec:double} we analyze under which conditions one may give rise to double gravitational layers and additional contributions to the stress-energy tensor of the hypersurface; in Sec. \ref{sec:strep} we describe how the junction conditions are affected if one introduces a dynamically equivalent scalar-tensor representation of the theory; and in Sec. \ref{sec:concl} we trace our conclusions.

\section{Theory and equations}\label{sec:theory}

In this work we are interested in studying modified theories of gravity that feature extra scalar degrees of freedom in comparison to GR, arising from arbitrary dependencies of the action in several scalar quantities. The action $S$ describing such theories can generally be written in the form
\begin{equation}\label{eq:action}
S=\frac{1}{2\kappa^2}\int_\Omega\sqrt{-g}\left[R+f\left(X_1,...,X_n\right)+2\kappa^2\mathcal L_m\right]d^4x,
\end{equation}
where $\kappa^2=8\pi G/c^4$, where $G$ is the gravitational constant and $c$ is the speed of light (in this work we shall adopt a system of geometrized units for which $G=c=1$ and thus $\kappa^2=8\pi$), $\Omega$ is the spacetime manifold written in terms of a set of coordinates $x^\mu$, $g$ is the determinant of the metric tensor $g_{\mu\nu}$, $R=g^{\mu\nu}R_{\mu\nu}$ is the Ricci scalar, with $R_{\mu\nu}$ the Ricci tensor, $f$ is a well-behaved function of several scalar quantities $X_i$ to be specified later, and $\mathcal L_m$ is the matter Lagrangian density.  

The modified field equations of these theories can be obtained via the application of the variational method to Eq.\eqref{eq:action} with respect to the metric $g_{\mu\nu}$. These equations take the general form
\begin{equation}\label{eq:field}
G_{\mu\nu}-\frac{1}{2}g_{\mu\nu}f\left(X_1,...,X_n\right)+\sum_{i=1}^n H_{\mu\nu}^{(i)}=8\pi T_{\mu\nu},
\end{equation}
where $G_{\mu\nu}=R_{\mu\nu}-\frac{1}{2}g_{\mu\nu}R$ is the Einstein's tensor, $T_{\mu\nu}$ is the matter stress-energy tensor defined in terms of the variation of the matter Lagrangian $\mathcal L_m$ in the usual way as
\begin{equation}\label{eq:def_Tab}
T_{\mu\nu}=-\frac{2}{\sqrt{-g}}\frac{\delta\left(\sqrt{-g}\mathcal L_m\right)}{\delta g^{\mu\nu}},
\end{equation}
and the tensors $H_{\mu\nu}^{(i)}$ correspond to the different contributions to the field equation of the dependency of the function $f$ in the scalar quantities $X_i$. These can be written explicitly in the form of a variation as
\begin{equation}\label{eq:def_Hab}
H_{\mu\nu}^{(i)}=\frac{\partial f}{\partial X_i}\frac{\delta X_i}{\delta g^{\mu\nu}}\equiv f_{X_i}\frac{\delta X_i}{\delta g^{\mu\nu}},
\end{equation}
where $f_{X_i}$ denotes the partial derivative of $f$ with respect to $X_i$. Given the linearity of Eq.\eqref{eq:field} in the terms $H_{\mu\nu}^{(i)}$, these quantities and the necessary conditions to preserve their regularity in the formalism of distribution functions can be analyzed independently.

In the literature, several different theories of gravity for which the action depends arbitrarily in some scalar quantity have been studied, including e.g. $f\left(R\right)$ theories, $f\left(T\right)$ theories where $T$ is the trace of the stress-energy tensor, among others. All of these theories fall into the family of actions given in Eq.\eqref{eq:action}, and their respective junction conditions can be analyzed following similar procedures. In this work, we take the quantities $X_i$ to belong to any subset of 
\begin{equation}\label{eq:def_scalars}
X_i=\{R,P,Q,\mathcal G,T,\mathcal T,\mathcal R\},
\end{equation}
where $R=g^{\mu\nu}R_{\mu\nu}$ and $T=g^{\mu\nu}T_{\mu\nu}$ are the previously defined Ricci scalar and trace of the stress-energy tensor $T_{\mu\nu}$ respectively, whereas the remaining scalars $P$, $Q$, $\mathcal G$, $\mathcal T$, and $\mathcal R$ are defined as follows
\begin{eqnarray}
&P=R_{\mu\nu}R^{\mu\nu},\\
&Q=R_{\mu\nu\sigma\rho}R^{\mu\nu\sigma\rho},\\
&\mathcal G=R^2-4P+Q,\\
&\mathcal T = T_{\mu\nu}T^{\mu\nu},\\
&\mathcal R = R_{\mu\nu}T^{\mu\nu}.
\end{eqnarray}
The quantity $\mathcal G$ is known as the Gauss-Bonnet invariant, with theories depending on $f\left(\mathcal G\right)$ being commonly referred to as Gauss-Bonnet theories of gravity. 

Regarding the theories of gravity featuring a dependency in a scalar that is proportional to the matter components, i.e., $T$, $\mathcal T$, and $\mathcal R$, although the junction conditions and the methods followed for their deduction are mostly unaltered, the field equations of these theories depend on the distribution of matter chosen, e.g. a vector field, a scalar field, or a fluid. For the purpose of this work, we take a relativistic isotropic perfect fluid as an example of application to outline the method. The stress-energy tensor and matter Lagrangian describing such a fluid are given by
\begin{equation}\label{eq:def_matter}
T_{\mu}^\nu=\text{diag}\left(-\rho,p,p,p\right), \qquad \mathcal L_m=p,
\end{equation}
where $\rho$ is the energy density and $p$ is the isotropic pressure. An extrapolation of the methods that follow to other types of matter distributions should be completely analogous.

\section{Distribution formalism}\label{sec:dist}

\subsection{Notation and assumptions}

Consider that the spacetime manifold $\Omega$ can be divided into two regions: an exterior region $\Omega^+$ described by a metric $g_{\mu\nu}^+$ written in terms of a coordinate system $x^\mu_+$, and an interior region $\Omega^-$ described by a metric $g_{\mu\nu}^-$ written in terms of a coordinate system $x^\mu_-$. These two regions intersect at a hypersurface $\Sigma$ on which one defines a set of coordinates $y^a$, where the latin indices exclude the direction perpendicular to $\Sigma$. Furthermore, we assume that there exists a coordinate system $x^\mu$ distinct from both $x^\mu_\pm$ that overlaps with these coordinate systems for some open sub-regions of $\Omega^\pm$ containing $\Sigma$, respectively\footnote{Note that this is an over-simplified approach to the problem. A more rigorous approach requires one to find an adequate diffeomorphism connecting the points in the two boundaries of $\Omega^\pm$, say $\Sigma^\pm$, thus unifying these two boundaries into a single matching hypersurface $\Sigma$, and to match the two tangent spaces at $\Sigma$. Nevertheless, once such an approach has been carried, the existence of a coordinate system valid in an open region that contains $\Sigma$ and continuous through $\Sigma$ is an acceptable assumption. For a more mathematically detailed approach, we refer the reader to \cite{Mars:1993mj}}. We define the projection vectors from the 4-dimensional spacetime into the 3-dimensional hypersurface as $e^\alpha_a=\partial x^\alpha/\partial y^a$, and a normal unit vector perpendicular to $\Sigma$ as $n^\mu$. The normal vector satisfies the normalization condition $n_\mu n^\mu=\epsilon$, with $\epsilon=\pm 1$ depending on $n^a$ being a timelike (negative sign) or spacelike (positive sign) vector. Furthermore, by construction one has $e^\alpha_a n_\alpha=0$, i.e., the projection of the normal vector into the hypersurface $\Sigma$ vanishes, as expected. The induced metric $h_{ab}^\pm$ and the extrinsic curvature $K_{ab}^\pm$ on each of the sides of $\Sigma$ can then be written as 
\begin{equation}\label{eq:def_induced}
h_{ab}^\pm=g_{\mu\nu}^\pm e^\mu_a e^\nu_b,
\end{equation}
\begin{equation}\label{eq:def_extrinsic}
K_{ab}^\pm=e^\mu_a e^\nu_b \nabla^\pm_\mu n_\nu,
\end{equation}
where $\nabla_\mu^\pm$ denotes a covariant derivative written in terms of the metrics $g_{\mu\nu}^\pm$. Consider a geodesic congruence orthogonal to the hypersurface $\Sigma$ and parametrized by an affine parameter $l$ such that different values of $l$ correspond to different hypersurfaces orthogonal to the geodesic congruence. Without loss of generality, one can define this parameter as positive for any hypersurface contained in $\Omega^+$, negative for any hypersurface contained $\Omega^-$, and zero at the hypersurface $\Sigma$, i.e., one can write $dx^\mu=n^\mu dl$ or, equivalently, $n_\mu=\epsilon \partial_\mu l$. We emphasize that different values of the parameter $l$ should not be interpreted as different points along a one-dimensional line, but instead to different three-dimensional hypersurfaces.

A suitable mathematical framework to conduct the analysis that follows is the distributional formalism. In this formalism, any regular quantity $X$ can be written in the form
\begin{equation}\label{eq:def_dist}
X=X^+\Theta\left(l\right)+X^-\Theta\left(-l\right),
\end{equation}
where the superscripts $X^\pm$ denote the quantity $X$ in the spacetime regions $\Omega^\pm$ respectively, and $\Theta\left(l\right)$ is the Heaviside distribution function, defined as $\Theta\left(l\right)=0$ for $l<0$, $\Theta\left(l\right)=1$ for $l>0$, and $\Theta\left(l\right)=\frac{1}{2}$ for $l=0$. In this formalism, one can also define the jump $\left[X\right]$ and the surface value $\left\{X\right\}$ of a quantity $X$ as
\begin{equation}\label{eq:def_jump}
\left[X\right]=X^+|_\Sigma-X^-|_\Sigma,
\end{equation}
\begin{equation}\label{eq:def_surface}
\left\{X\right\}=\frac{1}{2}\left(X^+|_\Sigma+X^-|_\Sigma\right).
\end{equation}
We note that there is a fundamental difference between the quantity $X$ defined in Eq. \eqref{eq:def_dist}, which is a quantity defined in the whole spacetime manifold $\Omega$ with two well-defined limits on $\Sigma$ with potentially different values $X^\pm|_\Sigma$, and quantities that are defined only on $\Sigma$, e.g. the quantities $h_{ab}^\pm$ and $K_{ab}^\pm$ defined in Eqs. \eqref{eq:def_induced} and \eqref{eq:def_extrinsic}, respectively, but with two potentially different values on each of the sides of $\Sigma$ e.g. $h^\pm_{ab}$ and $K^\pm_{ab}$. Although we define the jump and the surface values for both types quantities using the same notation (see Eqs. \eqref{eq:def_jump} and \eqref{eq:def_surface}), the quantities $X^\pm|_\Sigma$ should be interpreted as $X^\pm|_\Sigma=\lim_{l\to 0^\pm}X$, whereas the quantities $h^\pm_{ab}$ and $K^\pm_{ab}$ correspond to the values of these quantities on each side of $\Sigma$.  
These two definitions satisfy the following useful product properties
\begin{equation}\label{eq:def_prop}
\left[XY\right]=\left[X\right]\left\{Y\right\}+\left\{X\right\}\left[Y\right],
\end{equation}
\begin{equation}\label{eq:def_prop2}
\left\{XY\right\}=\left\{X\right\}\left\{Y\right\}+\frac{1}{4}\left[X\right]\left[Y\right],
\end{equation}
for any two quantities $X$ and $Y$. By definition, one has $\left[n^\mu\right]=0$ and $\left[e^\mu_a\right]=0$. Finally, taking the derivative of Eq.\eqref{eq:def_dist}, one verifies that the derivatives of a regular quantity $X$ can be written as
\begin{equation}\label{eq:def_ddist}
\nabla_\mu X=\nabla_\mu X^+\Theta\left(l\right)+\nabla_\mu X^-\Theta\left(-l\right)+\epsilon n_\mu\left[X\right]\delta\left(l\right),
\end{equation}
where $\delta\left(l\right)\equiv d\Theta/ dl$ is the Dirac-delta distribution function. To simplify the notation, in what follows we implement the following definition
\begin{equation}
X^\pm \equiv X^+\Theta\left(l\right)+X^-\Theta\left(-l\right),
\end{equation}
to represent the regular part of any quantity $X$.

\subsection{Quantities in the distribution formalism}

To analyze the junction conditions of a given theory of gravity, all relevant quantities must be expressed in the distribution formalism. While different theories of gravity may feature different relevant quantities, there are a few quantities e.g. the metric $g_{\mu\nu}$ and the Ricci scalar $R$ which are common among every theory under study. Thus, in this section we write these general quantities in the distribution formalism. We start by writing the metric $g_{ab}$ in the form
\begin{equation}\label{eq:def_metric}
g_{\mu\nu}=g_{\mu\nu}^+\Theta\left(l\right)+g_{\mu\nu}^-\Theta\left(-l\right)\equiv g_{\mu\nu}^\pm.
\end{equation}
To construct the Christoffel symbols $\Gamma^\alpha_{\mu\nu}$ associated with the metric $g_{\mu\nu}$, it is necessary to take the partial derivative of Eq.\eqref{eq:def_metric}. Following Eq.\eqref{eq:def_ddist}, one obtains $\partial_\alpha g_{\mu\nu}=\partial_\alpha g_{\mu\nu}^+\Theta\left(l\right)+\partial_\alpha g_{\mu\nu}^-\Theta\left(-l\right)+\epsilon n_\alpha\left[g_{\mu\nu}\right]\delta\left(l\right)$. Although the term proportional to $\delta\left(l\right)$ is not necessarily problematic at the level of the Christoffel symbols, when one tries to write the Riemann tensor $R_{\mu\nu\sigma\rho}$ in the distribution formalism, this quantity features products of Christoffel symbols, which consequently feature factors of the form $\delta^2\left(l\right)$. These factors are singular in the distribution formalism and must be eliminated to preserve the regularity of the Riemann tensor. To get rid of these problematic factors, one must impose the continuity of the metric, i.e., $\left[g_{\mu\nu}\right]=0$. Since $\left[e^\mu_a\right]=0$, this condition can be rewritten in a coordinate-invariant way by projecting both indices of the metric $g_{\mu\nu}$ into the hypersurface, from which one obtains the first junction condition
\begin{equation}\label{eq:junction_1}
\left[h_{ab}\right]=0.
\end{equation}
Notice how this junction condition arises as a straightforward consequence of the distribution formalism itself, and it is independent of the theory of gravity used as framework, as long as it is a metric theory of gravity. Following this result, the partial derivatives of $g_{\mu\nu}$ reduce to the regular form
\begin{equation}\label{eq:def_dmetric}
\partial_\alpha g_{\mu\nu}=\partial_\alpha g_{\mu\nu}^+\Theta\left(l\right)+\partial_\alpha g_{\mu\nu}^-\Theta\left(-l\right)\equiv \partial_\alpha g_{\mu\nu}^\pm.
\end{equation}

The Christoffel symbols associated with the metric $g_{ab}$ can now be constructed from Eqs.\eqref{eq:def_metric} and \eqref{eq:def_dmetric} and have a regular form. Consequently, one is now able to compute the form of the Riemann tensor $R_{\mu\nu\sigma\rho}$ in the distribution formalism, as well as its contractions, i.e., the Ricci tensor $R_{\mu\nu}=R^\sigma_{\ \mu\sigma\nu}$ and the Ricci scalar $R=g^{\mu\nu}R_{\mu\nu}$. These quantities take the forms
\begin{equation}\label{eq:dist_Rabcd}
R_{\mu\nu\sigma\rho}=R^\pm_{\mu\nu\sigma\rho}+\bar R_{\mu\nu\sigma\rho}\delta\left(l\right),
\end{equation}
\begin{equation}\label{eq:dist_Rab}
R_{\mu\nu}=R^\pm_{\mu\nu}+\bar R_{\mu\nu}\delta\left(l\right),
\end{equation}
\begin{equation}\label{eq:dist_R}
R=R^\pm+\bar R\delta\left(l\right),
\end{equation}
where the quantities $\bar R_{\mu\nu\sigma\rho}$, $\bar R_{\mu\nu}$ and $\bar R$ collectively denote the factors proportional to $\delta\left(l\right)$ and are given in terms of the fundamental geometrical quantities at $\Sigma$ as
\begin{equation}\label{eq:def_barRabcd}
\bar R_{\mu\nu\sigma\rho}=4\left[K_{ab}\right]e^a_{[\mu}n_{\nu]}e^b_{[\rho}n_{\sigma]},
\end{equation}
\begin{equation}\label{eq:def_barRab}
\bar R_{\mu\nu}=-\left(\epsilon\left[K_{ab}\right]e^a_\mu e^b_\nu+n_\mu n_\nu \left[K\right]\right),
\end{equation}
\begin{equation}\label{eq:def_barR}
\bar R=-2\epsilon\left[K\right],
\end{equation}
where we have defined index anti-symmetrization as $X_{[ab]}=\frac{1}{2}\left(X_{ab}-X_{ba}\right)$, and $K^\pm=h^{ab}_\pm K_{ab}^\pm$ is the trace of the extrinsic curvature. 

Regarding the matter sector, the terms proportional to $\delta\left(l\right)$ appearing in the field equations are associated with a thin-shell of matter present at the hypersurface $\Sigma$. To study the properties of this thin-shell, it is useful to write the stress-energy tensor $T_{\mu\nu}$ and its trace $T$ in the distribution formalism as\footnote{We note that the forms chosen for $T_{\mu\nu}$ and $T$ in Eqs. \eqref{eq:dist_tab} and \eqref{eq:dist_t}, respectively, are not the most general forms of these quantities. Indeed, these forms exclude more complicated scenarios for which contributions non-tangential to $\Sigma$ and double gravitational layers arise in the stress-energy tensor $T_{\mu\nu}$. Nevertheless, the forms taken are sufficient for the analysis of the junction conditions given in Secs. \ref{sec:juncgen} and \ref{sec:juncpart}. The analysis of the scenarios for which additional terms arise is undergone later in Sec. \ref{sec:double}, for which the corresponding form of the stress-energy tensor $T_{\mu\nu}$ is given in Eq. \eqref{eq:def_Tab_double}.}
\begin{equation}\label{eq:dist_tab}
T_{\mu\nu}=T_{\mu\nu}^\pm+S_{\mu\nu}\delta\left(l\right),
\end{equation}
\begin{equation}\label{eq:dist_t}
T=T^\pm+S\delta\left(l\right),
\end{equation}
where $S_{\mu\nu}=S_{ab}e^a_\mu e^b_\nu$, $S_{ab}$ is the 3-dimensional stress-energy tensor of the thin-shell, and $S=S_\mu^\mu$ is the trace of $S_{\mu\nu}$. The explicit form of $S_{ab}$ in terms of the geometrical quantities varies depending on the theory chosen as framework and it is determined explicitly by the junction conditions. In what follows, we define the contribution of each of the tensors $H_{\mu\nu}^{(i)}$ to the stress-energy tensor $S_{ab}$ by $S_{ab}^{(i)}$. In the end, the complete stress-energy tensor can be obtained from its individual contributions upon applying the relevant junction conditions for the combination of $X_i$ scalars considered.

\section{General set of junction conditions}\label{sec:juncgen}

The junction conditions are a set of conditions that the geometrical and matter quantities must satisfy in order to guarantee that two spacetime solutions defined in two complementary spacetime submanifolds of $\Omega$ can be matched to produce a regular solution of the field equations for the whole manifold, i.e., in both the two submanifolds $\Omega^+$ and $\Omega^-$, as well as the boundary hypersurface $\Sigma$. Since the solutions $g_{\mu\nu}^+$ and $g_{\mu\nu}^-$ are assumed to be solutions of the field equations for the regions $\Omega^+$ and $\Omega^-$ respectively, any possible incompatibilities appear precisely at $\Sigma$, upon matching the two solutions (hence the junction conditions are sometimes referred to as the matching conditions). 

To deduce the junction conditions of a given theory, one must proceed as follows. First, all the relevant quantities that appear in the field equations of that theory must be written in the distribution formalism. These quantities must be non-singular, i.e., they must not feature any terms proportional to $\delta^2\left(l\right)$. If any quantities in the field equation feature singular terms, these terms must be forced to vanish via the imposition of extra constraints. These constraints are precisely the junction conditions. When all singular terms have been successfully removed from the field equations, i.e., only terms proportional to $\Theta\left(l\right)$ and $\delta\left(l\right)$ remain, one can take the projection of the field equations into the hypersurface $\Sigma$ using the projection vectors $e^\mu_a e^\nu_b$. This allows one to obtain the explicit expression for the stress-energy tensor of the thin-shell $S_{ab}$. Finally, if one wants to require the smoothness of the matching, i.e., in the absence of a thin-shell, all of the terms proportional to $\delta\left(l\right)$ in the field equations must also be removed via the imposition of extra constraints, which add to the complete set of junction conditions.  

In this section, we analyze the contribution of each of the scalars identified in Eq.\eqref{eq:def_scalars} to the complete set of junction conditions of the theory. For this purpose, we discriminate the junction conditions arising from these contributions into three different subsets, as follows:
\begin{enumerate}
    \item \textbf{Immediate}: the junction conditions arising directly from writing the relevant geometrical and matter quantities and their respective derivatives in the distribution formalism and imposing their regularity;
    \item \textbf{Differential}: the junction conditions arising from differential terms depending on partial derivatives of the function $f\left(X_1,...,X_n\right)$ in the field equations that give rise to higher-order derivatives of the relevant quantities;
    \item \textbf{Coupling}: the junction conditions induced from a coupling of different scalars in the function $f\left(X_1,...,X_n\right)$, e.g. products $X_i X_j$ for $i\neq j$.
\end{enumerate}
Although the separation between immediate and differential junction conditions may appear quite arbitrary at this point, this distinction is relevant further on, upon analyzing the couplings between different scalars $X_i$. We start by analyzing the immediate and differential junction conditions arising from each of the scalars $X_i$ via their respective tensors $H_{\mu\nu}^{(i)}$, i.e., we analyze which conditions are necessary to maintain the regularity of the tensors $H_{\mu\nu}^{(i)}$ at the hypersurface $\Sigma$. This can be done by assuming that the function $f$ depends exclusively on the scalar $X_i$ under consideration, and allows one to extract the contributions $S_{ab}^{(i)}$ to the stress-energy tensor of the thin shell of each of the scalars $X_i$ independently. Afterwards, in Sec \ref{sec:couplings}, we analyze the coupling junction conditions arising from different combinations of scalars $X_i$ in the function $f$.

In what follows, we assume that $f$ is a general well-behaved function of its arguments, and that it admits a Taylor-series expansion, i.e., it is an analytic function. Note that particular forms of the theory for which certain partial derivatives vanish might lead to different scenarios, e.g. the rise of double gravitational layers. These special cases are analyzed afterwards in Secs. \ref{sec:juncpart} and \ref{sec:double}.

\subsection{Contribution of General Relativity}

Before jumping into the contributions of the different scalars $X_i$ to the junction conditions of the theory, let us first review the junction conditions of GR. Imposing $f=0$ in the field equations in Eq.\eqref{eq:field}, one recovers the Einstein's field equations $G_{\mu\nu}=8\pi T_{\mu\nu}$. Taking the distribution representations of $R_{\mu\nu}$, $R$ and $T_{\mu\nu}$ in Eqs.\eqref{eq:dist_Rab}, \eqref{eq:dist_R}, and \eqref{eq:dist_tab} respectively, and projecting the result into the hypersurface $\Sigma$ with $e_a^\mu e_b^\nu$, one obtains the contribution of the GR component of the theory to the stress-energy tensor of the thin-shell $S_{ab}^{GR}$ as
\begin{equation}\label{eq:GR_Sab}
8\pi S^{GR}_{ab}=-\epsilon\left(\left[K_{ab}\right]-\left[K\right]h_{ab}\right).
\end{equation}
This well-known result is the so-called second junction condition in GR, or Israel junction condition. Even though this result is only applicable in GR, as its validity comes directly from the fact that the field equations are the Einstein's field equations, several published works mistakenly use this junction condition to compute the stress-energy tensor of a thin-shell in modified theories of gravity described by different modified field equations, resulting in a consequent result lacking physical and mathematical significance. In the particular case of a smooth matching, i.e., in the absence of a thin-shell, or $S_{ab}=0$, the trace of Eq. \eqref{eq:GR_Sab} implies that $\left[K\right]=0$, which upon a replacement back into the original equation leads to the second junction condition of the form
\begin{equation}\label{eq:GR_jc_2}
\left[K_{ab}\right]=0,
\end{equation}
also known as the Darmois junction condition. In the following, we verify how these results extend to theories of gravity featuring an action with dependencies in the scalars $X_i$.

\subsection{Contribution of $f\left(R\right)$}

Let us start by considering the contributions of $f\left(R\right)$, where $R$ is the Ricci scalar. These junction conditions have already been analyzed in detail in Ref. \cite{senovilla1}, but here we provide a brief review to maintain the self consistency of the manuscript, as they are needed in what follows to construct more complicated theories e.g. $f\left(R,T\right)$ and $f\left(R,\mathcal G\right)$ gravity. For $X_i=R$, the tensor $H_{\mu\nu}^{(R)}$ takes the form
\begin{equation}\label{eq:def_H_R}
H_{\mu\nu}^{(R)}=f_R R_{\mu\nu}-\left(\nabla_\mu\nabla_\nu-g_{\mu\nu}\Box\right)f_R
\end{equation}
where $\Box=\nabla^\mu\nabla_\nu$ represents the d'Alembert operator. For a general analytic $f\left(R\right)$ function, i.e., that admits a Taylor-series expansion, one expects that powers of the form $R^2$ or higher generally appear in $f\left(R\right)$, as well as its partial derivatives $f_R$. Since $R$ features a term proportional to $\delta\left(l\right)$ in the distribution formalism, see Eq. \eqref{eq:dist_R}, these powers of $R$ feature terms proportional to $\delta^2\left(l\right)$ or higher, which are singular in the distribution formalism. To avoid these problematic terms and preserve the regularity of $f$ and its partial derivatives, it is necessary that the terms proportional to $\delta\left(l\right)$ in $R$ vanish, i.e., $\bar R=0$. Consequently, from Eq.\eqref{eq:def_barR}, one obtains an immediate junction condition in the form
\begin{equation}\label{eq:R_jc_1}
\left[K\right]=0.
\end{equation}

Consider now the differential terms in Eq.\eqref{eq:def_H_R}. Taking the first and second covariant derivatives of the function $f_R$ and using the chain rule, one is able to write these differential terms in the form of derivatives of $R$ as
\begin{equation}\label{eq:R_df}
\nabla_\mu f_R=f_{RR} \nabla_\mu R,
\end{equation}
\begin{equation}\label{eq:R_ddf}
\nabla_\mu\nabla_\nu f_R = f_{RRR} \nabla_\mu R\nabla_\nu R + f_{RR} \nabla_\mu\nabla_\nu R.
\end{equation}
The first-order covariant derivative $\nabla_\mu R$ can be obtained by taking a derivative of Eq.\eqref{eq:dist_R} subjected to the immediate junction condition in Eq.\eqref{eq:R_jc_1}, from which one obtains
\begin{equation}\label{eq:R_dR}
    \nabla_\mu R = \nabla_\mu R^\pm+\epsilon n_\mu\left[R\right]\delta\left(l\right),
\end{equation}
which features a term proportional to $\delta\left(l\right)$. Thus, the product $\nabla_\mu R\nabla_\nu R$ in Eq.\eqref{eq:R_ddf} gives rise to a singular $\delta^2\left(l\right)$ term in the second order derivatives of $f_R$. To prevent the appearance of this problematic term and preserve the regularity of $\nabla_\mu\nabla_\nu f_R$, one needs to impose the continuity of the Ricci scalar, i.e., we obtain the differential junction condition
\begin{equation}\label{eq:R_jc_2}
\left[R\right]=0.
\end{equation}

Finally, the second-order covariant derivative of $R$ can be obtained by taking a covariant derivative of Eq.\eqref{eq:R_dR} subjected to the junction condition in Eq.\eqref{eq:R_jc_2}, from which one obtains
\begin{equation}\label{eq:R_ddR}
\nabla_\mu\nabla_\nu R = \nabla_\mu\nabla_\nu R^\pm+\epsilon n_\mu\left[\nabla_\nu R\right]\delta\left(l\right).
\end{equation}

One can now deduce the contribution of the $f\left(R\right)$ component to the stress-energy tensor of the thin-shell, $S_{ab}^{(R)}$. Taking the field equations in Eq.\eqref{eq:field} with $H_{\mu\nu}$ given by Eq.\eqref{eq:def_H_R}, with the quantities in the distribution formalism given in Eqs.\eqref{eq:dist_Rab}, \eqref{eq:dist_R} and \eqref{eq:R_ddR}, subjected to the junction conditions in Eqs.\eqref{eq:R_jc_1} and \eqref{eq:R_jc_2}, and taking a projection into the hypersurface $\Sigma$ with $e_a^\mu e_b^\nu$, one obtains
\begin{equation}\label{eq:R_Sab}
8\pi S_{ab}^{(R)} = -\epsilon f_R\left[K_{ab}\right]+\epsilon h_{ab}f_{RR}n^\mu\left[\nabla_\mu R\right].
\end{equation}

Summarizing, for any theory of gravity featuring an arbitrary dependence $f\left(R\right)$ in its action, the complete set of junction conditions feature two extra junction conditions in comparison to GR, namely Eqs.\eqref{eq:R_jc_1} and \eqref{eq:R_jc_2}, and the stress-energy tensor of the thin-shell is given by Eq.\eqref{eq:R_Sab}. If one further requires that the matching is smooth, i.e., in the absence of a thin-shell, from Eq.\eqref{eq:R_Sab} one verifies that two additional junction conditions must be satisfied:
\begin{equation}
\left[K_{ab}\right]=0, \qquad \left[\nabla_\mu R\right]=0,
\end{equation}
which again features one extra condition with respect to GR.

\subsection{Contribution of $f\left(P\right)$}

Next, we consider the contributions of a function $f\left(P\right)$ where $P=R_{\mu\nu}R^{\mu\nu}$. In this situation, the tensor $H_{\mu\nu}^{(P)}$ takes the form
\begin{equation}\label{eq:def_H_P}
H_{\mu\nu}^{(P)}=2f_Pg^{\sigma\rho}R_{\mu\sigma}R_{\nu\rho}+g_{\mu\nu}\nabla_\sigma\nabla_\rho\left(f_P R^{\sigma\rho}\right)+\Box\left(f_P R_{\mu\nu}\right)-2\nabla_\sigma\nabla_\rho\left(f_P R^\sigma_{(\mu}\delta^\rho_{\nu)}\right),
\end{equation}
where we have defined index symmetrization as $X_{(ab)}=\frac{1}{2}\left(X_{ab}+X_{ba}\right)$. Since the Ricci tensor $R_{\mu\nu}$ features a term proportional to $\delta\left(l\right)$ in the distribution formalism, see Eq.\eqref{eq:dist_Rab}, this implies that, unlike the previous cases studied so far, the scalar $P$ itself features singular terms proportional to $\delta^2\left(l\right)$. In general, one can write $P$ in the distribution formalism as
\begin{equation}\label{eq:dist_P}
P=P^\pm+2\bar P \delta\left(l\right)+\hat P\delta^2\left(l\right),
\end{equation}
where the quantities $\bar P$ and $\hat P$ are given in terms of geometrical quantities at $\Sigma$ as
\begin{equation}\label{eq:def_barP}
\bar P = -\left\{R_{\mu\nu}\right\}\left(\epsilon\left[K^{ab}\right]e^{\mu}_ae^\nu_b+n^\mu n^\nu\left[K\right]\right),
\end{equation}
\begin{equation}\label{eq:def_hatP}
\hat P =\left[K_{ab}\right]\left[K^{ab}\right]+\left[K\right]^2.
\end{equation}
Since the two terms in $\hat P$ in Eq.\eqref{eq:def_hatP} are quadratic and, thus, strictly positive, the regularity of $P$ forcefully requires the extrinsic curvature to coincide on both sides of $\Sigma$, i.e., we obtain an immediate junction condition of the form
\begin{equation}\label{eq:P_jc_1}
\left[K_{ab}\right]=0.
\end{equation}
Consequently, both $\hat P$ and $\bar P$ vanish and $P$ remains regular. In the absence of any terms proportional to $\delta\left(l\right)$ in $P$, the regularity of $f\left(P\right)$ as well as its partial derivatives $f_P$ is guaranteed. Furthermore, another consequence of Eq.\eqref{eq:P_jc_1} is that $\bar R_{\mu\nu}=0$ also vanishes, thus preserving the regularity of the first term on the right-hand side of Eq.\eqref{eq:def_H_P}.

Consider now the differential terms in Eq.\eqref{eq:def_H_P}. Expanding these differential terms, one finds terms proportional to derivatives of the Ricci scalar $R_{\mu\nu}$, as well as terms proportional to the derivatives of the function $f_P$. The terms proportional to derivatives of $R_{\mu\nu}$ are associated with additional immediate junction conditions imposing the regularity of these derivatives, whereas the terms proportional to derivatives of $f_P$ are associated with the differential junction conditions. Let us start by analyzing the former. Taking the first-order derivative of Eq.\eqref{eq:dist_Rab} under the junction condition in Eq.\eqref{eq:P_jc_1}, one obtains
\begin{equation}\label{eq:P_dRab}
\nabla_\sigma R_{\mu\nu}=\nabla_\sigma R_{\mu\nu}^\pm+\epsilon n_\sigma \left[R_{\mu\nu}\right]\delta\left(l\right),
\end{equation}
Although the presence of the term proportional to $\delta\left(l\right)$ at this point is not problematic, since the tensor $H_{\mu\nu}$ does not feature products of first order derivatives of $R_{\mu\nu}$, its presence leads to the appearance of double gravitational layers, which are analyzed separately in the later Sec. \ref{sec:double}. Thus, to prevent these additional contributions to appear at this level, we impose the following immediate junction condition
\begin{equation}\label{eq:P_jc_3}
\left[R_{\mu\nu}\right]=0.
\end{equation}
Consequently, the second-order covariant derivative of $R_{\mu\nu}$ can be obtained by differentiating Eq.\eqref{eq:P_dRab} and taking Eq.\eqref{eq:P_jc_3} into consideration, from which one obtains
\begin{equation}
\nabla_\sigma\nabla_\rho R_{\mu\nu}=\nabla_\sigma\nabla_\rho R_{\mu\nu}^\pm+\epsilon n_\sigma \left[\nabla_\rho R_{\mu\nu}\right]\delta\left(l\right).
\end{equation}

Let us now turn to the terms with derivatives of the function $f_P$ in Eq. \eqref{eq:def_H_P}. Similarly to what happens in the $f\left(R\right)$ case, the first and second-order covariant derivatives of the function $f_P$ can be written in terms of derivatives of $P$ in the form
\begin{equation}\label{eq:P_df}
\nabla_\mu f_P=f_{PP} \nabla_\mu P,
\end{equation}
\begin{equation}\label{eq:P_ddf}
\nabla_\mu\nabla_\nu f_P = f_{PPP} \nabla_\mu P\nabla_\nu P + f_{PP} \nabla_\mu\nabla_\nu P.
\end{equation}
The first-order covariant derivative of $P$ can be obtained by taking a derivative of Eq.\eqref{eq:dist_P} subjected to the junction condition in Eq.\eqref{eq:P_jc_1}, which takes the form
\begin{equation}\label{eq:dist_dP}
\nabla_\mu P=\nabla_\mu P^\pm+\epsilon n_\mu \left[P\right]\delta\left(l\right).
\end{equation}
This derivative features a term proportional to $\delta\left(l\right)$, which gives rise to singular terms $\delta^2\left(l\right)$ in the first term of Eq.\eqref{eq:P_ddf} due to the products $\nabla_\mu P\nabla_\nu P$. To preserve the regularity of $\nabla_\mu\nabla_\nu f_P$, one must thus impose the continuity of $P$, i.e., we obtain a differential junction condition of the form\footnote{We note that even though $P=R_{\mu\nu}R^{\mu\nu}$, having $\left[P\right]=0$ does not imply immediately that $\left[R_{\mu\nu}\right]=0$. Indeed, if $\left\{R_{\mu\nu}\right\}=0$, it is possible that $\left[P\right]=0$ even though $\left[R_{\mu\nu}\right]\neq 0$, see Eq.\eqref{eq:def_prop}. However, the inverse is true, as having $\left[R_{\mu\nu}\right]=0$ implies $\left[P\right]=0$. Thus, this junction condition could be alternatively derived as a consequence of Eq. \eqref{eq:P_jc_3}.}
\begin{equation}\label{eq:P_jc_2}
\left[P\right]=0.
\end{equation}
Consequently, the second-order covariant derivative of $P$ can be obtained by differentiating Eq.\eqref{eq:dist_dP} subjected to the junction condition in Eq.\eqref{eq:P_jc_2}, which yields
\begin{equation}\label{eq:dist_ddp}
\nabla_\mu\nabla_\nu P=\nabla_\mu\nabla_\nu P^\pm+\epsilon n_\mu \left[\nabla_\nu P\right]\delta\left(l\right).
\end{equation}

We now have all the necessary ingredients to deduce the form of the stress-energy of the thin-shell $S_{ab}^{(P)}$. Taking the field equations in Eq.\eqref{eq:field}, with the distribution quantities in Eqs.\eqref{eq:dist_Rab}, \eqref{eq:dist_P}, and \eqref{eq:dist_tab}, subjected to the junction conditions in Eqs.\eqref{eq:P_jc_1}, \eqref{eq:P_jc_3}, and \eqref{eq:P_jc_2}, and taking a projection into $\Sigma$ with $e_a^\mu e_b^\nu$, one obtains
\begin{eqnarray}\label{eq:P_Sab}
8\pi S_{ab}^{(P)}&=&\epsilon h_{ab}\left(f_{PP}\left\{R^{\mu\nu}\right\}n_\mu\left[\nabla_\nu P\right]+f_p n_\mu\left[\nabla_\nu R^{\mu\nu}\right]\right)+ \\
&+&\epsilon e_a^\mu e_b^\nu\left(f_{PP}\left\{R_{\mu\nu}\right\}n^\sigma\left[\nabla_\sigma P\right]+f_P n^\sigma\left[\nabla_\sigma R_{\mu\nu}\right]-2f_{PP}\left\{R^\rho_{(\mu}\right\}n_{\nu)}\left[\nabla_\rho P\right]-2f_P n_{\rho}\left[\nabla_{(\mu} R^{\rho}_{\nu)}\right]\right).\nonumber
\end{eqnarray}

Summarizing, a theory of gravity featuring a dependency $f\left(P\right)$ in the action has a complete set of junction conditions with three extra conditions in comparison to GR, namely Eqs.\eqref{eq:P_jc_1}, \eqref{eq:P_jc_3}, and \eqref{eq:P_jc_2}, and the stress-energy tensor of the thin-shell is given by Eq.\eqref{eq:P_Sab}. Note that Eq.\eqref{eq:P_jc_1} also appears in GR but for the particular case of smooth matching, whereas in $f\left(P\right)$ gravity it is a necessary requirement even for a matching with a thin-shell. If one now requires a smooth matching, i.e., in the absence of a thin-shell, two extra junction conditions arise:
\begin{equation}
\left[\nabla_\mu P\right]=0, \qquad \left[\nabla_\sigma R_{\mu\nu}\right]=0,
\end{equation}
none of which are a requirement of GR.

\subsection{Contribution of $f\left(Q\right)$}

Let us now analyze the contributions of a function $f\left(Q\right)$, with $Q=R_{\mu\nu\sigma\rho}R^{\mu\nu\sigma\rho}$. For such a function, the tensor $H_{\mu\nu}^{(Q)}$ is given by
\begin{equation}\label{eq:def_H_Q}
H_{\mu\nu}^{(Q)}=2f_QR_{\sigma\rho\gamma\mu}R^{\ \sigma\rho\gamma}_\nu-4\nabla_\sigma\nabla_\rho\left(f_Q {R^\sigma_{\ (\mu\nu)}}^\rho\right).
\end{equation}
Similarly to what happens for the scalar $P$, since the Riemann tensor features a term proportional to $\delta\left(l\right)$ in the distribution formalism, see Eq. \eqref{eq:dist_Rabcd}, then the scalar $Q$ features singular terms proportional to $\delta^2\left(l\right)$. In general, the scalar $Q$ can be written in the distribution formalism in the form
\begin{equation}\label{eq:dist_Q}
Q=Q^\pm+2\bar Q\delta\left(l\right)+\hat Q\delta^2\left(l\right),
\end{equation}
where the quantities $\bar Q$ and $\hat Q$ are given in terms of geometrical quantities at $\Sigma$ as
\begin{equation}
\bar Q = 4\left\{R_{\mu\nu\sigma\rho}\right\}\left[K_{ab}\right]e^a_{[\mu}n_{\nu]}e^b_{[\rho}n_{\sigma]},
\end{equation}
\begin{equation}
\hat Q=4\left[K_{ab}\right]\left[K^{ab}\right].
\end{equation}
Since $\hat Q$ is quadratic in $\left[K_{ab}\right]$ and, thus, strictly positive, to preserve the regularity of $Q$ it is necessary to force the extrinsic curvature to coincide on both sides of $\Sigma$, i.e., one obtains an immediate junction condition of the form
\begin{equation}\label{eq:Q_jc_1}
    \left[K_{ab}\right]=0.
\end{equation}
Following this result, both the quantities $\bar Q$ and $\hat Q$ vanish and $Q$ becomes regular. Since now $Q$ does not feature terms proportional to $\delta\left(l\right)$, the function $f\left(Q\right)$ and its partial derivative $f_Q$ are automatically regular. Note also that Eq.\eqref{eq:Q_jc_1} implies that $\bar R_{\mu\nu\sigma\rho}=0$, thus guaranteeing the regularity of the first term on the right-hand side of Eq.\eqref{eq:def_H_Q}.

Concerning the differential term in Eq.\eqref{eq:def_H_Q}, which gives rise to one term proportional to derivatives of ${R^\sigma_{\ (\mu\nu)}}^\rho$ and one term proportional to derivatives of the function $f_Q$, let us start by analyzing the former. Taking the covariant derivative of Eq.\eqref{eq:dist_Rabcd} and taking the junction condition in Eq.\eqref{eq:Q_jc_1} into consideration, one obtains
\begin{equation}\label{eq:dist_dRabcd}
\nabla_\gamma R_{\mu\nu\sigma\rho} = \nabla_\gamma R_{\mu\nu\sigma\rho}^\pm + \epsilon n_\gamma \left[R_{\mu\nu\sigma\rho} \right]\delta\left(l\right).
\end{equation}
Although the term proportional to $\delta\left(l\right)$ in the equation above is not problematic by itself, since the tensor $H_{\mu\nu}$ does not feature any products of first-order derivatives of $R_{\mu\nu\sigma\rho}$, its presence leads to the appearance of double gravitational layers, a topic that we deal with in detail in an upcoming section, Sec.\ref{sec:double}. Thus, to prevent the double gravitational layers to rise at this point, we impose the immediate junction condition
\begin{equation}\label{eq:Q_jc_3}
\left[R_{\mu\nu\sigma\rho}\right]=0.
\end{equation}
The second order covariant derivatives of $R_{\mu\nu\sigma\rho}$ can now be obtained by differentiating Eq.\eqref{eq:dist_dRabcd} and imposing Eq.\eqref{eq:Q_jc_3}, from which we obtain
\begin{equation}
\nabla_\gamma\nabla_\delta R_{\mu\nu\sigma\rho}=\nabla_\gamma\nabla_\delta R_{\mu\nu\sigma\rho}^\pm + \epsilon n_\gamma\left[\nabla_\delta R_{\mu\nu\sigma\rho}\right]\delta\left(l\right).
\end{equation}

Let us now turn to the term proportional to the derivative of $f_Q$ in Eq. \eqref{eq:def_H_Q}. The first and second-order covariant derivatives of the function $f_Q$ can be written in terms of derivatives of $Q$ via the chain rule, which take the forms
\begin{equation}\label{eq:Q_df}
\nabla_\mu f_Q=f_{QQ} \nabla_\mu Q,
\end{equation}
\begin{equation}\label{eq:Q_ddf}
\nabla_\mu\nabla_\nu f_Q = f_{QQQ} \nabla_\mu Q\nabla_\nu Q + f_{QQ} \nabla_\mu\nabla_\nu Q.
\end{equation}
Taking a derivative of Eq.\eqref{eq:dist_Q} subjected to the junction condition in Eq.\eqref{eq:Q_jc_1}, the first-order covariant derivative of $Q$ takes the form
\begin{equation}\label{eq:dist_dQ}
\nabla_\mu Q=\nabla_\mu Q^\pm+\epsilon n_\mu\left[Q\right] \delta\left(l\right).
\end{equation}
Since this derivative features a term proportional to $\delta\left(l\right)$, it gives rise to singular terms $\delta^2\left(l\right)$ in the first term of Eq.\eqref{eq:Q_ddf} due to the products $\nabla_\mu Q\nabla_\nu Q$, and thus the continuity of $Q$ is necessary to preserve the regularity of this term, i.e., one obtains a differential junction condition of the form\footnote{Again, note that even though $Q=R_{\mu\nu\sigma\rho}R^{\mu\nu\sigma\rho}$, having $\left[Q\right]=0$ does not imply immediately that $\left[R_{\mu\nu\sigma\rho}\right]=0$, as the same result can be obtained from $\left\{R_{\mu\nu\sigma\rho}\right\}=0$, see Eq.\eqref{eq:def_prop}. However, the inverse is true, which implies that one could have deduced this junction condition alternatively as a consequence of Eq. \eqref{eq:Q_jc_3}.}
\begin{equation}\label{eq:Q_jc_2}
\left[Q\right]=0.
\end{equation}
The second-order covariant derivative of $Q$ can now be obtained by differentiating Eq.\eqref{eq:dist_dQ} subjected to the junction condition in Eq.\eqref{eq:Q_jc_2}, from which one obtains
\begin{equation}
\nabla_\mu\nabla_\nu Q=\nabla_\mu\nabla_\nu Q^\pm+\epsilon n_\mu\left[\nabla_\nu Q\right]\delta\left(l\right).
\end{equation}

We are now equipped with all the necessary quantities to obtain the explicit form of the contribution $S_{ab}^{(Q)}$ to the stress-energy tensor of the thin-shell. Taking the field equations in Eq.\eqref{eq:field} with the distribution functions in Eq.\eqref{eq:dist_Rabcd}, \eqref{eq:dist_Q}, and \eqref{eq:dist_t}, along with the junction conditions in Eqs.\eqref{eq:Q_jc_1}, \eqref{eq:Q_jc_3}, and \eqref{eq:Q_jc_2}, and projecting the result into $\sigma$ with $e^\mu_a e^\nu_b$, we obtain
\begin{equation}\label{eq:Q_Sab}
S_{ab}^{(Q)}=-4\epsilon n_\sigma\left(f_{QQ}\left[\nabla_\rho Q\right]\left\{{R^\sigma_{\ (\mu\nu)}}^\rho\right\}+f_Q\left[\nabla_\rho {R^\sigma_{\ (\mu\nu)}}^\rho\right]\right).
\end{equation}

Summarizing, for a theory of gravity with an action depending on $f\left(Q\right)$, three extra junction conditions arise in comparison with GR, namely Eqs.\eqref{eq:Q_jc_1}, \eqref{eq:Q_jc_3}, and \eqref{eq:Q_jc_2}, and the contributions of this dependency to the stress-energy tensor of the thin-shell are given in Eq.\eqref{eq:Q_Sab}. Similarly to the $f\left(P\right)$ case, the condition in Eq.\eqref{eq:Q_jc_1}, which in GR appears only for smooth matching, here it is a requirement even for the matching with a thin-shell. To require a smooth matching, i.e., without a thin-shell, two extra junction conditions must be imposed,
\begin{equation}
\left[\nabla_\mu Q\right]=0,\qquad \left[\nabla_\gamma R_{\mu\nu\sigma\rho}\right]=0,
\end{equation}
again none of which are a requirement in GR.

\subsection{Contribution of $f\left(\mathcal G\right)$}

Consider now the contributions of a function $f\left(\mathcal G\right)$, where $\mathcal G$ is the Gauss-Bonnet invariant. The tensor $H_{\mu\nu}^{(\mathcal G)}$ in this situation takes the form

\begin{eqnarray}\label{eq:def_H_G}
H_{\mu\nu}^{(\mathcal G)}=\frac{1}{2}g_{\mu\nu}\mathcal G f_{\mathcal G}+4\left(R_{\mu \sigma\nu \rho}+2g_{\sigma[\nu}R_{\rho]\mu}+2g_{\mu[\sigma}G_{\nu]\rho}\right)\nabla^\sigma\nabla^\rho f_\mathcal{G}.
\end{eqnarray}
Since the Riemann tensor and its contractions feature terms proportional to $\delta\left(l\right)$, see Eqs.\eqref{eq:dist_Rabcd} to \eqref{eq:dist_R}, then the scalar $\mathcal G$ could in principle feature terms proportional to $\delta^2\left(l\right)$. Even though this is not the case, as we prove in what follows, let us for now write $\mathcal G$ in the distribution formalism in the form
\begin{equation}\label{eq:dist_G}
\mathcal G = \mathcal G^\pm+2\bar{\mathcal G}\delta\left(l\right)+\hat{\mathcal G}\delta^2\left(l\right),
\end{equation}
where the quantities $\bar{\mathcal G}$ and $\hat{\mathcal G}$ are given in terms of the geometrical quantities at $\Sigma$ as
\begin{equation}\label{eq:def_barG}
\bar{\mathcal G}=\left\{R\right\}\bar R-4\left\{R_{ab}\right\}\bar R^{ab}+\left\{R_{abcd}\right\}\bar R^{abcd},
\end{equation}
\begin{equation}\label{eq:def_hatG}
\hat{\mathcal G}=\bar R^2-4\bar R_{ab}\bar R^{ab}+\bar R_{abcd}\bar R^{abcd}.
\end{equation}
The term proportional to $\delta^2\left(l\right)$ is singular in the distributional formalism and thus should be removed, similarly to what was done in the previous cases. However, inserting Eqs.\eqref{eq:def_barRabcd} to \eqref{eq:def_barR} into Eq.\eqref{eq:def_hatG}, one verifies that the terms in the right-hand side of the latter equation cancel out, resulting in $\hat{\mathcal G}=0$. Thus, no extra junction conditions arise from this potentially singular term. On the other hand, replacing Eqs.\eqref{eq:def_barRabcd} to \eqref{eq:def_barR} into Eq.\eqref{eq:def_barG}, an adequate algebraic manipulation allows one to rewrite $\bar{\mathcal G}$ in the convenient form
\begin{equation}
\bar{\mathcal G}=4\epsilon\left\{R_{abcd}\right\}h^{\gamma\delta}e^a_\gamma e^c_\delta e^b_\alpha e^d_\beta\left(\left[K^{\alpha\beta}\right]-\frac{1}{2}h^{\alpha\beta}\left[K\right]\right).
\end{equation}
For a general analytical function $f\left(\mathcal G\right)$, one expects that powers of the form $\mathcal G^2$ or higher should appear in $f\left(\mathcal G\right)$ and its partial derivatives $f_\mathcal{G}$. Given that $\mathcal G$ still features a term proportional to $\delta\left(l\right)$, these powers of $\mathcal G$ would give rise to singular $\delta^2\left(l\right)$ factors in the distribution formalism. To preserve the regularity of $f$ and its partial derivatives, one must thus impose $\bar{\mathcal G}=0$, i.e., $\left[K_{ab}\right]=\frac{1}{2}h_{ab}\left[K\right]$. Taking the trace of this condition, one obtains $\left[K\right]=0$, which upon a replacement back in the original constraint leads to the immediate junction condition of the form
\begin{equation}\label{eq:G_jc_1}
\left[K_{ab}\right]=0
\end{equation}

Regarding now the differential terms in Eq.\eqref{eq:def_H_G}, the first and second order partial derivatives of $f_\mathcal{G}$ can be written in terms of derivatives of $\mathcal G$ as
\begin{equation}\label{eq:G_df}
\nabla_\mu f_\mathcal{G}=f_{\mathcal{G}\mathcal{G}} \nabla_\mu \mathcal{G},
\end{equation}
\begin{equation}\label{eq:G_ddf}
\nabla_\mu\nabla_\nu f_\mathcal{G} = f_{\mathcal{G}\mathcal{G}\mathcal{G}} \nabla_\mu \mathcal{G}\nabla_\nu \mathcal{G} + f_{\mathcal{G}\mathcal{G}} \nabla_\mu\nabla_\nu \mathcal{G}.
\end{equation}
Similarly to the previous cases, the first-order covariant derivative of $\mathcal G$ can be obtained by taking a derivative of Eq.\eqref{eq:dist_G} under the junction condition in Eq.\eqref{eq:G_jc_1}, from which one obtains
\begin{equation}\label{eq:dist_dG}
\nabla_\mu\mathcal{G}=\nabla_\mu\mathcal{G}^\pm+\epsilon n_\mu\left[\mathcal{G}\right]\delta\left(l\right).
\end{equation}
This derivative features a term proportional to $\delta^2\left(l\right)$ which, upon a replacement into Eq.\eqref{eq:G_ddf}, gives rise to singular terms $\delta^2\left(l\right)$ due to the products $\nabla_\mu\mathcal{G}\nabla_\nu\mathcal{G}$. To preserve the regularity of $\nabla_\mu\nabla_\nu f_\mathcal{G}$, the continuity of $\mathcal{G}$ must be imposed, i.e., we obtain the differential junction condition
\begin{equation}\label{eq:G_jc_2}
\left[\mathcal{G}\right]=0.
\end{equation}
The second-order covariant derivative of $\mathcal{G}$ can then be obtain by taking a derivative of Eq.\eqref{eq:dist_dG} subjected to the condition in Eq.\eqref{eq:G_jc_2}, which takes the form
\begin{equation}\label{eq:dist_ddG}
\nabla_\mu\nabla_\nu\mathcal{G}=\nabla_\mu\nabla_\nu\mathcal{G}^\pm+\epsilon n_\mu\left[\nabla_\nu\mathcal{G}\right]\delta\left(l\right).
\end{equation}
One now has all the necessary ingredients to obtain the contribution of the $f\left(\mathcal G\right)$ component to the stress-energy tensor of the thin-shell, $S_{ab}^{(\mathcal G)}$. Taking the field equations in Eq.\eqref{eq:field} with a $H_{\mu\nu}$ tensor given by Eq.\eqref{eq:def_H_G}, with the quantities in the distribution formalism given in Eq.\eqref{eq:dist_Rabcd}, \eqref{eq:dist_Rab}, \eqref{eq:dist_R}, \eqref{eq:dist_ddG}, and considering the junction conditions in Eqs.\eqref{eq:G_jc_1} and \eqref{eq:G_jc_2}, and projecting the result into $\Sigma$ using $e^\mu_a e^\nu_b$, one obtains
\begin{equation}\label{eq:G_Sab}
8\pi S_{ab}^{(\mathcal{G})}=4\epsilon e^\mu_a e^\nu_b g^{\gamma\rho}\left[\nabla_\gamma\mathcal G\right]n^\sigma f_{\mathcal G\mathcal G} \left( \left\{R_{\mu\sigma\nu\rho}\right\}+2g_{\sigma[\nu}\left\{R_{\rho]\mu}\right\}+2g_{\mu[\sigma}\left\{G_{\nu]\rho}\right\}\right).
\end{equation}
Summarizing, for a theory of gravity with an action depending on $f\left(\mathcal G\right)$, two extra junction conditions arise in comparison with GR, namely Eqs.\eqref{eq:G_jc_1} and \eqref{eq:G_jc_2}, and the contributions of this function to the stress-energy tensor of the thin-shell are given by Eq.\eqref{eq:G_Sab}. Similarly to what happens in $f\left(P\right)$ and $\left(Q\right)$, the condition in Eq.\eqref{eq:G_jc_1} appears even if the matching is not smooth, unlike in GR for which it is only necessary if the matching is smooth. To require now a smooth matching, i.e., in the absence of a thin-shell, a single extra junction condition must be imposed, namely
\begin{equation}
\left[\nabla_\mu\mathcal G\right]=0,
\end{equation}
a condition that is not required in GR.

\subsection{Contribution of $f\left(T\right)$}

Let us now consider the contributions of a function $f\left(T\right)$. These junction conditions have already been indirectly deduced in Ref.\cite{rosafrt} for the $f\left(R,T\right)$ gravity, but again we briefly review them here for self consistency. For $X_i=T$, the tensor $H_{\mu\nu}^{(T)}$ takes the form
\begin{equation}\label{eq:def_H_T}
H_{\mu\nu}^{(T)}=f_T\left(T_{\mu\nu}+\Theta_{\mu\nu}\right),
\end{equation}
where $\Theta_{\mu\nu}$ is an auxiliary tensor defined in terms of the variation of the stress-energy tensor $T_{\mu\nu}$ as
\begin{equation}
\Theta_{\mu\nu}=g^{\sigma\rho}\frac{\delta T_{\sigma\rho}}{\delta g^{\mu\nu}}=-2T_{\mu\nu}+g_{\mu\nu}\mathcal L_m-2g^{\sigma\rho}\frac{\partial \mathcal L_m}{\partial g^{\mu\nu}\partial g^{\sigma\rho}}.
\end{equation}
For the matter distribution specified in Eq.\eqref{eq:def_matter}, the auxiliary tensor $\Theta_{\mu\nu}$ takes the form
\begin{equation}
\Theta_{\mu\nu}=-2T_{\mu\nu}+pg_{\mu\nu}.
\end{equation}
For a function $f\left(T\right)$ that admits a Taylor-series expansion, one expects that in general powers of the form $T^2$ or higher should appear both in $f\left(T\right)$ and its partial derivatives $f_T$. From Eq.\eqref{eq:dist_t}, one verifies that $T$ features a term proportional to $\delta\left(l\right)$, which induces terms $\delta^2\left(l\right)$ in the function $f\left(T\right)$, which are singular in the distribution formalism. To preserve the regularity of $f\left(T\right)$ and its derivative $f_T$, one must thus impose that the terms proportional to $\delta\left(l\right)$ in $T$ must vanish, i.e., we obtain the immediate junction condition
\begin{equation}\label{eq:T_jc_1}
S=0.
\end{equation}
Since there are no differential terms in Eq.\eqref{eq:def_H_T}, the $F\left(T\right)$ theory does not feature any differential junction conditions, and one can proceed directly to the analysis of the contribution of this theory to the stress-energy tensor of the thin-shell $S_{\mu\nu}^{(T)}$. Taking the field equations in Eq.\eqref{eq:field} with $H_{\mu\nu}$ given by Eq.\eqref{eq:def_H_T}, the stress-energy tensor $T_{\mu\nu}$ written in the distribution formalism from Eq. \eqref{eq:dist_tab}, and taking a projection into $\Sigma$ with $e_a^\mu e_b^\nu$, one obtains
\begin{equation}\label{eq:T_Sab}
8\pi S_{ab}^{(T)}=f_T S_{ab}.
\end{equation}
Note that the two tensors $S_{ab}^{(T)}$ and $S_{ab}$ represent different quantities: the first represents the single contribution of the $f\left(T\right)$ component to the thin-shell, whereas the second represents the complete stress-energy tensor of the thin-shell. The junction condition in Eq.\eqref{eq:T_jc_1} that requires $S=0$ concerns only the complete stress-energy tensor, i.e., the quantity $S_{ab}$, and must be implemented only after all contributions of the $X_i$ scalars have been introduced.

Summarizing, for a theory of gravity featuring a dependency $f\left(T\right)$ in the action, the complete set of junction conditions features one extra condition in comparison to GR, namely Eq.\eqref{eq:T_jc_1}, and the contribution to the stress-energy tensor of the thin-shell is given by Eq.\eqref{eq:T_Sab}. Requiring the matching to be smooth, i.e., in the absence of a thin-shell or $S_{ab}=0$, does not give rise to any additional junction conditions at this point. Similarly to the condition $S=0$, as the necessary requirements to achieve smoothness depend on the explicit form of $S_{ab}$, one can only impose smoothness after the complete theory to be used as framework has been specified.

\subsection{Contribution of $f\left(\mathcal T\right)$}

Consider now the contributions of a function $f\left(\mathcal T\right)$, where $\mathcal T = T_{\mu\nu}T^{\mu\nu}$. In this situation, the tensor $H_{\mu\nu}^{(\mathcal T)}$ takes the form
\begin{equation}\label{eq:def_H_T2}
    H_{\mu\nu}^{(\mathcal T)}=f_\mathcal{T}\Phi_{\mu\nu},
\end{equation}
where $\Phi_{\mu\nu}$ is an auxiliary tensor defined in terms of the variation of the scalar $\mathcal T$ as
\begin{equation}
\Phi_{\mu\nu}=\frac{\delta \mathcal T}{\delta g^{\mu\nu}}=-2\mathcal L_m\left(T_{\mu\nu}-\frac{1}{2}g_{\mu\nu}T\right)-T T_{\mu\nu}+2T_{\mu}^\sigma T_{\sigma\nu}-4T^{\sigma\rho}\frac{\partial^2\mathcal L_m}{\partial g^{\mu\nu}g^{\sigma\rho}}.
\end{equation}
For the matter content specified in Eq.\eqref{eq:def_matter}, the auxiliary tensor $\Phi_{\mu\nu}$ takes the form
\begin{equation}
\Phi_{\mu\nu}=-2p\left(T_{\mu\nu}-\frac{1}{2}g_{\mu\nu}T\right)-T T_{\mu\nu}+2T_\mu^\sigma T_{\sigma\nu}.
\end{equation}
The stress-energy tensor $T_{\mu\nu}$ features a term proportional to $\delta\left(l\right)$, namely the stress-energy tensor of a possible thin-shell at $\Sigma$, see Eq.\eqref{eq:dist_tab}. This implies that the scalar $\mathcal T$ features singular terms proportional to $\delta^2\left(l\right)$ in its distributional form. More precisely, the scalar $\mathcal T$ can be written in the general form
\begin{equation}\label{eq:dist_T2}
\mathcal T=\mathcal T^\pm+2\bar{\mathcal T}\delta\left(l\right)+\hat{\mathcal T}\delta^2\left(l\right),
\end{equation}
where the quantities $\bar{\mathcal T}$ and $\hat{\mathcal T}$ are given in terms of the stress-energy of the thin-shell as
\begin{equation}
\bar{\mathcal T}=\left\{T_{\mu\nu}\right\}S^{\mu\nu},
\end{equation}
\begin{equation}
\hat{\mathcal T}=S_{\mu\nu}S^{\mu\nu}.
\end{equation}
Since $\hat{\mathcal T}$ is quadratic in $S_{\mu\nu}$, and thus it is strictly positive, one verifies that the regularity of $\mathcal T$ is only preserved if no thin-shell is present upon matching, i.e., the immediate junction condition takes the form
\begin{equation}\label{eq:T2_jc_1}
S_{\mu\nu}=0.
\end{equation}
This is a highly restrictive condition, as if forces any matching in a theory for which the action depends on a function $f\left(\mathcal T\right)$ to be smooth. Following this result, both $\bar{\mathcal T}$ and $\hat{\mathcal T}$ vanish, and the scalar $\mathcal T$ is regular, as well as the function $f$. given that Eq.\eqref{eq:def_H_T2} does not feature any differential terms, no other junction conditions arise in this theory. Furthermore, since the junction condition in Eq.\eqref{eq:T2_jc_1} forces the matching to be smooth, the contribution of $f\left(\mathcal T\right)$ to the stress-energy tensor of the thin-shell $S_{ab}^{(\mathcal T)}$ vanishes identically, i.e.,
\begin{equation}
S_{ab}^{(\mathcal T)}=0.
\end{equation}

\subsection{Contribution of $f\left(\mathcal R\right)$}

Consider next the contributions of a function $f\left(\mathcal R\right)$, where $\mathcal R=R_{\mu\nu}T^{\mu\nu}$. For such a function, the tensor $H_{\mu\nu}^{(\mathcal R)}$ becomes
\begin{equation}\label{eq:def_H_RT}
    H_{\mu\nu}^{(\mathcal R)}=\frac{1}{2}\left(\Box T_{\mu\nu}f_\mathcal{R} +g_{\mu\nu} \nabla_\sigma\nabla_\rho T^{\sigma\rho}f_\mathcal{R}\right)-\nabla_\sigma\nabla_{(\mu}T^\sigma_{\nu)}f_\mathcal{R}-\Xi_{\mu\nu}f_\mathcal{R},
\end{equation}
where $\Xi_{\mu\nu}$ is an auxiliary tensor defined as
\begin{equation}
\Xi_{\mu\nu}=-G_{\mu\nu}\mathcal L_m-\frac{1}{2}RT_{\mu\nu}+2R^\sigma_\mu T_{\sigma\nu}-2R^{\sigma\rho}\frac{\delta^2\mathcal L_m}{\delta g^{\mu\nu}\delta g^{\sigma\rho}}.
\end{equation}
For the matter distribution considered in Eq.\eqref{eq:def_matter}, the auxiliary tensor $\Xi_{\mu\nu}$ takes the explicit form
\begin{equation}
\Xi_{\mu\nu}=-G_{\mu\nu}p-\frac{1}{2}R T_{\mu\nu}+2R_{(mu}^\sigma T_{\nu)\sigma}.
\end{equation}
Both the Ricci tensor $R_{\mu\nu}$ and the stress-energy tensor $T_{\mu\nu}$ feature terms proportional to $\delta\left(l\right)$ in the distribution formalism, see Eqs.\eqref{eq:dist_Rab} and \eqref{eq:dist_tab}, respectively. Thus, the scalar $\mathcal R$ features singular terms proportional to $\delta^2\left(l\right)$. The scalar $\mathcal R$ can be written in the general form
\begin{equation}\label{eq:dist_RT}
\mathcal R=\mathcal R^\pm+2\bar{\mathcal R}\delta\left(l\right)+\hat{\mathcal R}\delta^2\left(l\right),
\end{equation}
where the quantities $\bar{\mathcal R}$ and $\hat{\mathcal R}$ can be written in terms of both geometrical quantities and the stress-energy tensor of the thin-shell at $\Sigma$ as
\begin{equation}\label{eq:def_barRT}
\bar{\mathcal R}=\left\{R_{\mu\nu}\right\}S^{\mu\nu}-\left\{T_{\mu\nu}\right\}\left(\epsilon\left[K^{ab}\right]e^\mu_ae^\nu_b+n^\mu n^\nu\left[K\right]\right),
\end{equation}
\begin{equation}\label{eq:def_hatRT}
\hat{\mathcal R}=-S_{\mu\nu}\left(\epsilon\left[K^{ab}\right]e^\mu_ae^\nu_b+n^\mu n^\nu\left[K\right]\right).
\end{equation}
To preserve the regularity of $\mathcal R$, the quantity $\hat{\mathcal R}$ must be forced to vanish. This can be achieved either via the imposition $S_{\mu\nu}=0$ or $\left[K_{ab}\right]=0$. However, even though either of this assumptions guarantees the regularity of $\mathcal R$, the regularity of $f\left(\mathcal R\right)$ is not guaranteed until both of the assumptions are independently imposed. Indeed, since the function $f\left(\mathcal R\right)$ is assumed to be analytical, and thus can feature power-laws of the form $\mathcal R^2$ or higher, the presence of the term proportional to $\delta\left(l\right)$ in $\mathcal R$ could lead to singular terms $\delta^2\left(l\right)$ in $f\left(\mathcal R\right)$. To avoid these problematic terms, both $\hat{\mathcal R}$ and $\bar{\mathcal R}$ must be forced to vanish. If one forces $\hat{\mathcal R}$ to vanish by imposing $S_{\mu\nu}=0$, then $\bar{\mathcal R}$ can only vanish by imposing $\left[K_{ab}\right]=0$ and vice-versa. Consequently, two immediate junction conditions arising form the regularity of $f\left(\mathcal R\right)$ are
\begin{equation}\label{eq:RT_jc_1}
S_{\mu\nu}=0,
\end{equation}
\begin{equation}\label{eq:RT_jc_2}
\left[K_{ab}\right]=0.
\end{equation}
Again, these conditions are highly restrictive and force any matching in a theory of gravity for which the action depends in $f\left(\mathcal R\right)$ to be smooth. 

Consider now the differential terms in Eq.\eqref{eq:def_H_RT}. These contain first and second-order covariant derivatives of both $f_\mathcal{R}$ and $T_{\mu\nu}$, which we should analyze separately. Taking the covariant derivative of Eq. \eqref{eq:dist_tab} under the condition of smoothness obtained in Eq. \eqref{eq:RT_jc_1}, one obtains
\begin{equation}\label{eq:dist_dTab}
\nabla_\sigma T^{\mu\nu}=\nabla_\sigma T^{\mu\nu}_\pm + \epsilon n_\sigma\left[T^{\mu\nu}\right]\delta\left(l\right).
\end{equation}
The term proportional to $\delta\left(l\right)$ would is not problematic by itself. However, its presence would lead to derivatives of the $\delta\left(l\right)$ function which are associated with double gravitational layers, which are studied later in Sec. \ref{sec:double}. Thus, to avoid the analysis of double gravitational layers at this point, we impose the immediate junction condition
\begin{equation}\label{eq:RT_jc_5}
\left[T^{\mu\nu}\right]=0.
\end{equation}
The second order covariant derivatives of $T^{\mu\nu}$ can now be obtained by differentiating Eq. \eqref{eq:dist_dTab} and imposing Eq. \eqref{eq:RT_jc_5}, from which one obtains
\begin{equation}
    \nabla_\sigma\nabla_\rho T^{\mu\nu}=\nabla_\sigma\nabla_\rho T^{\mu\nu}_\pm+\epsilon n_\sigma \left[\nabla_\rho T^{\mu\nu}\right]\delta\left(l\right).
\end{equation}

Let us now turn to the terms proportional to derivatives of $f_\mathcal{R}$ in Eq. \eqref{eq:def_H_RT}. The derivatives of $f_\mathcal{R}$ can be written in terms of derivatives of $\mathcal R$ as
\begin{equation}\label{eq:RT_df}
\nabla_\mu f_\mathcal{R}=f_{\mathcal{R}\mathcal{R}} \nabla_\mu \mathcal{R},
\end{equation}
\begin{equation}\label{eq:RT_ddf}
\nabla_\mu\nabla_\nu f_\mathcal{R} = f_{\mathcal{R}\mathcal{R}\mathcal{R}} \nabla_\mu \mathcal{R}\nabla_\nu \mathcal{R} + f_{\mathcal{R}\mathcal{R}} \nabla_\mu\nabla_\nu \mathcal{R}.
\end{equation}
The first-order derivative of $\mathcal R$ can be obtained by taking a derivative of Eq.\eqref{eq:dist_RT} subjected to the junction conditions in Eqs.\eqref{eq:RT_jc_1} and \eqref{eq:RT_jc_2}, which yields
\begin{equation}\label{eq:dist_dRT}
\nabla_\mu\mathcal R=\nabla_\mu\mathcal R^\pm+\epsilon n_\mu\left[\mathcal R\right]\delta\left(l\right).
\end{equation}
The derivative above features a term proportional to $\delta\left(l\right)$ which, when replaced into Eq.\eqref{eq:RT_ddf}, leads to singular products of the form $\delta^2\left(l\right)$ due to the products $\nabla_\mu\mathcal R\nabla_\nu\mathcal R$. Thus, to preserve the regularity of $\nabla_\mu\nabla_\nu f_\mathcal{R}$, one must impose the continuity of $\mathcal R$, i.e., we obtain the differential junction condition
\begin{equation}\label{eq:RT_jc_3}
\left[\mathcal R\right]=0.
\end{equation}
Then, the second-order derivative of $\mathcal R$ can be obtained by taking a derivative of Eq.\eqref{eq:dist_dRT} under the condition in Eq.\eqref{eq:RT_jc_3}, from which one obtains
\begin{equation}
\nabla_\mu\nabla_\nu\mathcal R=\nabla_\mu\nabla_\nu\mathcal R^\pm+\epsilon n_\nu\left[\nabla_\mu\mathcal R\right]\delta\left(l\right).
\end{equation}
One can now insert these results back into the definition of $H_{\mu\nu}^{(\mathcal R)}$ given in Eq.\eqref{eq:def_H_RT}, take a projection into $\Sigma$ using $e_a^\mu e_b^\nu$ and verify what terms proportional to $\delta\left(l\right)$ appear in this tensor. Note that, since in this case the matching is forced to be smooth from the junction condition in Eq.\eqref{eq:RT_jc_1}, these terms are not contributions to the stress-energy tensor of the thin-shell $S_{ab}^{(\mathcal R)}$. Indeed, this tensor vanishes identically, i.e., 
\begin{equation}
S_{ab}^{(\mathcal R)}=0.
\end{equation}
Instead, the appearance of extra terms proportional to $\delta\left(l\right)$ in $H_{\mu\nu}^{(\mathcal R)}$ give rise to extra junction conditions. In particular, the first-order derivative of $\mathcal R$ and $T^{\mu\nu}$ must be continuous, i.e.,
\begin{equation}\label{eq:RT_jc_4}
\left[\nabla_\mu\mathcal R\right]=0,
\end{equation}
\begin{equation}\label{eq:RT_jc_6}
\left[\nabla_\sigma T^{\mu\nu}\right]=0.
\end{equation}
Note that, since the first of these conditions arises from a term proportional to a covariant derivative of $f_\mathcal{R}$, it corresponds to a differential junction condition, whereas the second one, which comes from a term proportional to a covariant derivative of $T^{\mu\nu}$, corresponds to an immediate junction condition. This distinction is important at the level of the coupling junction conditions, as is shown in Sec. \ref{sec:couplings}.

Summarizing, for a theory of gravity for which the action depends explicitly in $f\left(\mathcal R\right)$, the matching is always forced to be smooth, i.e., $S_{ab}=0$, and four extra junction conditions arise in comparison with GR, namely the ones provided in Eqs.\eqref{eq:RT_jc_1}, \eqref{eq:RT_jc_2}, \eqref{eq:RT_jc_5}, \eqref{eq:RT_jc_3}, \eqref{eq:RT_jc_4}, and  \eqref{eq:RT_jc_6}.

\subsection{Contribution of couplings $X_i X_j$}\label{sec:couplings}

In the previous sections, we have derived the junction conditions of a certain $f\left(X\right)$ theory of gravity assuming that the action function depends on a single scalar $X$. If one considers a theory depending on two or more scalars, say $f\left(X_i,X_j\right)$, additional differential junction conditions arising from the coupling of the two scalars may emerge, as well as extra contributions to the stress-energy tensor of the thin-shell, when it is present. To illustrate how this happens, let us consider a theory of gravity for which action depends on a function of two scalars $X$ and $Y$, i.e., $f\left(X,Y\right)$. For such a  theory of gravity, the first and second-order covariant derivatives take the forms
\begin{equation}\label{eq:Cdf}
\nabla_\mu f = f_X\nabla_\mu X + f_Y\nabla_\mu Y,
\end{equation}
\begin{equation}\label{eq:Cddf}
\nabla_\mu\nabla_\nu f=f_{XX}\nabla_\mu X\nabla_\nu X+f_{YY}\nabla_\mu Y\nabla_\nu Y+2f_{XY}\nabla_{(\mu}X\nabla_{\nu)}Y+f_X\nabla_\mu\nabla_\nu X+f_Y\nabla_\mu\nabla_\nu Y.
\end{equation}

Consider that the scalars $X$ and $Y$ are such that the tensor $H_{\mu\nu}^X$ of $f\left(X\right)$ features at least one differential term (this happens for the scalars $R$, $P$, $Q$, $\mathcal G$, and $\mathcal R$), say e.g. $\nabla_\mu\nabla_\nu f_X$, whereas the tensor $H_{\mu\nu}^Y$ of $f\left(Y\right)$ does not feature any differential terms (this happens for the scalars $T$ and $\mathcal T$). In this case, the previous analysis of $f\left(X\right)$ and $f\left(Y\right)$ would lead to the appearance of differential junction conditions for $X$ but not for $Y$. However, if the function $f\left(X,Y\right)$ depends simultaneously in both $X$ and $Y$, Eq. \eqref{eq:Cddf} tells us that the terms that induce the differential junction conditions for the scalar $X$ have exact counterparts for the scalar $Y$, in this case $\nabla_\mu\nabla_\nu Y$, even if the field equations for $f\left(Y\right)$ only do not feature such terms. This implies that, when $X$ and $Y$ are coupled in a theory $f\left(X,Y\right)$, this theory features differential junction conditions for the field $Y$ with the same form as the ones for the field $X$. 

Under the same assumptions as the previous paragraph, an analysis of the $f\left(X\right)$ theory would have led to a contribution to the stress-energy tensor of the thin-shell $S_{ab}$ proportional to the second-order derivative of $X$, whereas the analysis of $f\left(Y\right)$ would not have led to such a contribution. Again, if the function $f\left(X,Y\right)$ depends simultaneously on both $X$ and $Y$, Eq. \eqref{eq:Cddf} shows that the term proportional to $f_X\nabla_\mu\nabla_\nu X$ which induces such a contribution to $S_{ab}$ has a counterpart proportional to $f_Y\nabla_\mu\nabla_\nu Y$. Thus, even if the function $f\left(Y\right)$ by itself does not induce a contribution to the stress-energy tensor $S_{ab}$, the coupling between $X$ and $Y$ in $f\left(X,Y\right)$ does induce such a contribution, of the same form as the one induced by $f\left(X\right)$ alone.

Summarizing the statements traced in the previous two paragraphs, for any theory of gravity for which the action depends on more than one scalar $X_i$, the couplings between any two scalars $X_i$ and $X_j$ imply that:
\begin{enumerate}
    \item The coupling junction conditions for $X_j$ induced by the coupling with $X_i$ are of the same form as the differential junction conditions for $X_i$ but applied to $X_j$;
    \item The additional contributions of $X_j$ to the stress-energy tensor $S_{ab}$ induced by the coupling with $X_i$ are of the same form as the contributions of $X_i$ to the stress-energy tensor $S_{ab}$ arising from the differential terms in $H_{\mu\nu}^{X_i}$, where the factor proportional to $f_{X_i X_i}$ is exchanged by a factor $f_{X_i X_j}$.
\end{enumerate}

\subsubsection{Example of coupling conditions: $f\left(R,T\right)$ gravity}

To clarify the procedure described above, let us provide an explicit example on how to compute the contributions of the couplings between different scalars to the junction conditions. Consider a theory for which the action is described by a function $f\left(R,T\right)$. According to the results of the previous sections, a function $f\left(R\right)$ induces an immediate junction condition $\left[K\right]=0$, a differential junction condition $\left[R\right]=0$, and the differential terms contribute to the stress-energy tensor with a term proportional to $f_{RR}n^\mu\left[\nabla_\mu R\right]$. On the other hand, the function $f\left(T\right)$ induces an immediate junction condition $S=0$, and due to the absence of differential terms in $H_{\mu\nu}^{(T)}$ it does not induce any differential junction condition. Now, point number $1$ above states that the theory $f\left(R,T\right)$ should feature an additional coupling junction condition of the same for as the differential junction condition in $f\left(R\right)$ but applies to $T$, i.e., one obtains the coupling junction condition $\left[T\right]=0$. Furthermore, point number $2$ above states that the theory $f\left(R,T\right)$ should also feature an additional contribution to the stress-energy tensor $S_{ab}$ of the same for as the contribution of the differential terms of $f\left(R\right)$ but applied to $T$, i.e., one obtains an additional contribution proportional to $f_{RT}n^\mu\left[\nabla_\mu T\right]$. The junction conditions for $f\left(R,T\right)$ gravity thus take the form:
\begin{eqnarray}
&\left[K\right]=0,\nonumber \\
&\left[R\right]=0,\nonumber \\
&\left[T\right]=0, \\
&S=0, \nonumber 
\end{eqnarray}
while the contributions to the stress-energy tensor of the thin-shell $S_{ab}^{(RT)}$ take the forms
\begin{equation}\label{eq:RTSab}
\left(8\pi+f_T\right) S_{ab}^{(RT)}=-\epsilon f_R\left[K_{ab}\right]+
\epsilon h_{ab} n^\mu\left(f_{RR}\left[\nabla_\mu R\right]+f_{RT}\left[\nabla_\mu T\right]\right).
\end{equation}
Now, taking the trace of Eq.\eqref{eq:RTSab} and using the junction conditions $S=0$ and $\left[K\right]=0$, one obtains $f_{RR}\left[\nabla_\mu R\right]+f_{RT}\left[\nabla_\mu T\right]=0$, which can then be replaced again into Eq.\eqref{eq:RTSab} to simplify the result. The final set of junction conditions for the $f\left(R,T\right)$ theory of gravity thus takes the form
\begin{eqnarray}
&\left[K\right]=0,\nonumber \\
&\left[R\right]=0,\nonumber \\
&\left[T\right]=0, \\
&f_{RR}\left[\nabla_\mu R\right]+f_{RT}\left[\nabla_\mu T\right]=0, \nonumber \\
&\left(8\pi+f_T\right) S_{ab}=-\epsilon f_R\left[K_{ab}\right],\nonumber
\end{eqnarray}
which corresponds precisely to the set of junction conditions previously found in Ref.\cite{rosafrt}. The same procedure can be applied to any other theory with an action defined by a function $f\left(X_1,...,X_n\right)$.

\subsection{Summary of junction conditions}

Let us now summarize the results obtained in this section. The immediate and differential junction conditions for the matching between two spacetimes arising from the dependency of the action of the theory in each of the scalars analyzed, as well as the additional junction conditions that arise if one further requires the matching to be smooth, are summarized in Table \ref{tab:conditions}. Note that in case one is interested in analyzing a theory featuring more than one scalar, i.e., with couplings between different scalars, the additional coupling junction conditions and contributions to the stress-energy tensor of the thin-shell can be extracted following the procedure outlined in Sec. \ref{sec:couplings}.

\begin{table}
\begin{tabular}{c|c c c}
    Theory & Immediate & Differential & Smooth\\ \hline \\
    GR & None & None & $\left[K_{ab}\right]=0$ \\ \\
    $f\left(R\right)$ & $\left[K\right]=0$ & $\left[R\right]=0$ & $\begin{matrix} \left[K_{ab}\right]=0 \\ \left[\nabla_\mu R\right]=0 \end{matrix}$ \\ \\
    $f\left(P\right)$ & $\begin{matrix}\left[K_{ab}\right]=0 \\ \left[R_{\mu\nu}\right]=0\end{matrix}$ & $\left[P\right]=0$ & $\begin{matrix} \left[\nabla_\mu P\right]=0 \\ \left[\nabla_\sigma R_{\mu\nu}\right]=0 \end{matrix}$ \\ \\
    $f\left(Q\right)$ & $\begin{matrix}\left[K_{ab}\right]=0 \\ \left[R_{\mu\nu\sigma\rho}\right]=0\end{matrix}$ & $\left[Q\right]=0$  & $\begin{matrix} \left[\nabla_\mu Q\right]=0 \\ \left[\nabla_\gamma R_{\mu\nu\sigma\rho}\right]=0 \end{matrix}$ \\ \\
    $f\left(T\right)$ & $S=0$ & None & None \\ \\
    $f\left(\mathcal T\right)$ & $S_{\mu\nu}=0$ & None & Always \\ \\
    $f\left(\mathcal R\right)$ & $\begin{matrix}S_{\mu\nu}=0 \\ \left[K_{ab}\right]=0 \\ \left[T^{\mu\nu}\right]=0 \\ \left[\nabla_\sigma T^{\mu\nu}\right]=0\end{matrix}$ & $\begin{matrix}\left[\mathcal R\right]=0 \\ \left[\nabla_\mu\mathcal R\right]=0\end{matrix}$ & Always \\ 
\end{tabular}
\caption{Summary of the immediate junction conditions, differential junction conditions, and additional junction conditions required for smooth matching for each of the theories under analysis. For the additional junction conditions arising from couplings between different scalar, refer to Sec. \ref{sec:couplings}.}
\label{tab:conditions}
\end{table}

\section{Particular sets of junction conditions}\label{sec:juncpart}

In Sec. \ref{sec:juncgen} we outlined a general method to obtain the junction conditions of a theory of gravity described by an action depending on a general function $f\left(X_1,...,X_n\right)$ of scalars $X_i$. In this method, we have assumed that this function is as general as possible, i.e., it admits a Taylor-series expansion with non-vanishing coefficients up to any arbitrary expansion order. If this assumption is dropped, i.e., if some specific coefficients of the Taylor-series expansion vanish, it might happen that some of the requirements of the full set of junction conditions and contributions to the stress-energy tensor of the thin-shell for a general theory may be discarded or are altered. Thus, in this section we analyze under which assumptions the full set of junction conditions is altered to a simpler form.

\subsection{Eliminate coupling junction conditions}

In Sec. \ref{sec:couplings}, we proved that additional junction conditions arise when the function $f\left(X_1,...,X_n\right)$ features couplings between the scalars $X_i$ and $X_j$, for $i\neq j$, if either (or both) of the scalars $X_i$ and $X_j$ induce differential junction conditions in the particular cases $f\left(X_i\right)$ and $f\left(X_j\right)$, which happens for the scalars $R$, $P$, $Q$, $\mathcal G$, and $\mathcal R$. Let us now analyze how particular forms of the function $f\left(X_1,...,X_n\right)$ may lead to a removal of any additional coupling junction conditions from the system.

Consider a scalar $X$ for which $H_{\mu\nu}^{(X)}$ features differential terms of the form $\nabla_\mu\nabla_\nu f_X$, that induce differential junction conditions in the particular case $f\left(X\right)$, coupled to another scalar $Y$. Following Eq.\eqref{eq:Cddf}, one verifies that the coupling junction conditions for $Y$ arise from terms in $\nabla_\mu\nabla_\nu f_X$ proportional to $f_{XYY}$ and $f_{XXY}$, whereas the additional contributions to the stress energy tensor $S_{ab}$ arise from terms proportional to $f_{XY}$. Thus, if one chooses a particular form of the function $f\left(X,Y\right)$ such that  $f_{XYY}=f_{XXY}=f_{XY}=0$, the resulting theory does not feature any additional coupling junction conditions, even though the action depends on both the scalars $X$ and $Y$. On the other hand, one could instead consider a function satisfying $f_{XYY}=f_{XXY}=0$ and $f_{XY}\neq 0$, from which one would obtain a theory featuring additional coupling junction conditions from $X$ to $Y$ but no additional contributions to the stress-energy tensor of the thin-shell. In general, for a theory depending on several scalar quantities, one can always select specific forms of the function $f\left(X_1,...,X_n\right)$ for which the terms proportional to partial derivatives that induce coupling junction conditions or additional contributions to the stress-energy tensor of the thin-shell caused by the coupling between any two specific scalars $X_i$ and $X_j$ are absent, while the couplings between any remaining scalars are present.  

The simplest possible way to avoid coupling junction conditions is to split the function $f\left(X_1,...X_n\right)$ into $n$ independent functions proportional to each of the scalars $X_i$ independently, i.e., 
\begin{equation}\label{eq:f_part}
f\left(X_1,...,X_n\right)=\sum_{i=1}^n f^{(i)}\left(X_i\right).
\end{equation}
Under this assumption, all crossed derivatives of the function $f$ vanish, and thus no additional coupling terms arise in the field equations. However, this is not the most general form that the function $f$ can have. Indeed, for some of the scalars $X_i$, the function $f$ may admit some specific low-order couplings without giving rise to coupling junction conditions. Thus, if one aims to use the most general function $f\left(X_1,...,X_n\right)$ that avoids the appearance of coupling junction conditions, the additional admissible terms in Eq.\eqref{eq:f_part} must be analyzed in a case-by-case basis, via the expansion of the differential terms associated to each of the scalars $X_i$ and identifying which partial derivatives of the function $f$ give rise to additional conditions.

As an example, consider again the function $f\left(R,T\right)$. The differential terms in $H_{\mu\nu}^{(R)}$ depend on second-order derivatives of the function $f_R$, which by Eq. \eqref{eq:Cddf} lead to
\begin{equation}\label{eq:df_part_RT}
\nabla_\mu\nabla_\nu f_R=f_{RRR}\nabla_\mu R\nabla_\nu R+f_{RTT}\nabla_\mu T\nabla_\nu T
+2f_{RRT}\nabla_{(\mu}R\nabla_{\nu)}T+f_{RR}\nabla_\mu\nabla_\nu R+f_{RT}\nabla_\mu\nabla_\nu T.
\end{equation}
One can verify directly that a function of the form of Eq. \eqref{eq:f_part}, i.e., $f\left(R,T\right)=f_1\left(R\right)+f_2\left(T\right)$, would lead to the vanishing of all coupling terms in the result above, thus preventing the appearance of additional coupling junction conditions. However, although this drastic split is sufficient, it is not necessary. Indeed, one verifies that the terms $\nabla_\mu T\nabla_\nu T$ and $\nabla_{(\mu}R\nabla_{\nu)}T$ that give rise to additional coupling conditions appear proportionally to the factors $f_{RTT}$ and $f_{RRT}$. These two factors vanish even if the function $f\left(R,T\right)$ features a product $RT$. Thus, the most general form of the function $f\left(R,T\right)$ that allows one to discard the coupling junction conditions from the full set of junction conditions is
\begin{equation}\label{eq:f_part_RT}
f\left(R,T\right)=f_1\left(R\right)+f_2\left(T\right)+\alpha RT,
\end{equation}
where $\alpha$ is a coupling constant. 

If one further demands that the additional contributions to the stress-energy tensor of the thin-shell caused by the couplings between $R$ and $T$ are absent, a similar analysis as before must be undergone but with respect to the differential term $\nabla_\mu\nabla_\nu T$. This term, which gives rise to additional contributions to $S_{ab}$, appears proportionally to a factor $f_{RT}$, which must be forced to vanish. In this case, the only possible way to achieve this result is to consider $\alpha=0$ in Eq. \eqref{eq:f_part_RT}, thus recovering the complete split between $R$ and $T$. Under these assumptions, the complete set of junction conditions for the function $f\left(R,T\right)=f_1\left(R\right)+f_2\left(T\right)$ takes the form
\begin{eqnarray}
&\left[K\right]=0, \nonumber \\
&\left[R\right]=0,  \\
&\left[\nabla_\mu R\right]=0,\nonumber \\
&\left(8\pi+f_T\right) S_{ab}=-\epsilon f_R\left[K_{ab}\right]. \nonumber
\end{eqnarray}

\subsection{Eliminate differential junction conditions}
\label{sec:eliminatediff}

The procedure to eliminate the differential junction conditions is similar to the one outlined in the previous section to eliminate the coupling junction conditions. In this case, one must again expand the derivatives in the differential terms of $H_{\mu\nu}^{(i)}$, if any, and select a particular form of the function $f$ such that any terms giving rise to differential junction conditions in the field equations vanish. 

Taking again the function $f\left(R,T\right)$ as an example, the second-order derivatives of $f_R$ that appear in $H_{\mu\nu}^{(R)}$ are given in Eq.\eqref{eq:df_part_RT}. In this equation, the differential junction conditions of $f\left(R\right)$ arise from the terms $f_{RRR}\nabla_\mu R\nabla_\nu R$ and $f_{RRT}\nabla_\mu R\nabla_\nu T$, whereas the contribution to the stress-energy tensor of the thin-shell comes from the term $f_{RR}\nabla_\mu\nabla_\nu R$. Thus, if one searches a specific form of the theory for which the differential junction condition is discarded, one must select a form of $f\left(R,T\right)$ such that $f_{RRR}=0$ and $f_{RRT}=0$. The most general form of $f\left(R,T\right)$ that satisfies these requirements is
\begin{equation}
f\left(R,T\right)=\alpha R f_1\left(T\right) + \beta R^2 + f_2\left(T\right)
\end{equation}
where $\alpha$ and $\beta$ are coupling constants. For this choice of function $f$, the full set of junction conditions takes the form
\begin{eqnarray}
&\left[K\right]=0,\nonumber \\
&\left[T\right]=0,\nonumber \\
&f_{RR}\left[\nabla_\mu R\right]+f_{RT}\left[\nabla_\mu T\right]=0, \\
&\left(8\pi +f_T\right)S_{ab}=-\epsilon f_R\left[K_{ab}\right]. \nonumber
\end{eqnarray}
On the other hand, if one further requires that the contribution to the stress-energy tensor of the thin-shell is removed, then one must select a form of $f\left(R,T\right)$ such that $f_{RR}=0$, i.e., $f\left(R,T\right)$ must be at most linear in $R$. The most general function $f\left(R,T\right)$ that satisfies this requirement is
\begin{equation}
f\left(R,T\right)=R f_1\left(T\right)+f_2\left(T\right),
\end{equation}
where $\alpha$ and $\beta$ are coupling constants. For this choice of function $f$, the full set of junction conditions takes the form
\begin{eqnarray}
&\left[K\right]=0,\nonumber \\
&\left[T\right]=0,\nonumber \\
&\left[\nabla_\mu T\right]=0, \\
&\left(8\pi +f_T\right)S_{ab}=-\epsilon f_R\left[K_{ab}\right]. \nonumber
\end{eqnarray}

Finally, if one demands that both the coupling and the differential junction conditions are discarded, then the function $f$ must satisfy all of the previous requirements simultaneously, i.e., $f_{RRR}=0$, $f_{RRT}=0$, $f_{RTT}=0$ $f_{RR}=0$, and $f_{RT}$=0, the first three being automatically covered by the latter two. One thus obtains a function $f$ of the form
\begin{equation}
f\left(R,T\right) = \alpha R + f\left(T\right).
\end{equation}
The full set of junction conditions thus reduces to
\begin{eqnarray}
    &\left[K\right]=0,  \\
    &\left(8\pi +f_T\right)S_{ab}=-\epsilon f_R\left[K_{ab}\right].\nonumber
\end{eqnarray}

\subsection{Eliminate immediate junction conditions}

In Sec. \ref{sec:juncgen}, we showed that the immediate junction conditions may arise in two different ways. If a scalar $X_i$ features a term proportional to $\delta^2\left(l\right)$ in their definition in the distribution formalism, the vanishing of this singular term induces an immediate junction condition. Any immediate junction condition arising from such a situation can not be discarded independently of the form of the function $f\left(X_1,...,X_n\right)$, as it comes from a problematic definition of the scalar $X_i$ itself. Examples of such conditions can be found e.g. for the scalars $P$, $Q$, $\mathcal T$, and $\mathcal R$. On the other hand, an immediate junction condition may arise for a scalar $X_i$ if it features a term proportional to $\delta\left(l\right)$ but no term proportional to $\delta^2\left(l\right)$ in their distribution formalism. In such a case, the immediate junction condition arises from the assumption that the function $f$ admits a Taylor-series expansion and thus it features arbitrary powers of $X_i$, which induce singular powers of $\delta\left(l\right)$. Immediate junction conditions arising from the latter situation can thus be discarded by selecting a function $f\left(X_i,...,X_j\right)$ that does not feature any products of $\delta\left(l\right)$ functions. In other words, this can be achieved by forcing the function $f\left(X_i,...,X_j\right)$ to be linear and uncoupled in the scalar $X_i$ whose immediate junction conditions we want to remove from the system.

Taking again $f\left(R,T\right)$ as an example, which features the two immediate junction conditions $\left[K\right]=0$ from $f\left(R\right)$ and $S=0$ from $f\left(T\right)$, one thus verifies that the first of these conditions can be discarded by considering a function that is linear and uncoupled in $R$, i.e., by choosing the form
\begin{equation}
f\left(R,T\right)=\alpha R + f\left(T\right),
\end{equation}
whereas the second of these conditions can be discarded by considering a function that is linear and uncoupled in $T$, i.e., of the form
\begin{equation}
f\left(R,T\right)=f\left(R\right)+\alpha T,
\end{equation}
where $\alpha$ is a coupling constant. If one wants to discard both of the conditions above, then one must consider a function that is linear and uncoupled in both the scalars $R$ and $T$, that is
\begin{equation}
f\left(R,T\right)=\alpha R + \beta T,
\end{equation}
where $\alpha$ and $\beta$ are free parameters. Note that if the function was to depend on more than two scalars, say e.g. $R$, $T$, and $\mathcal G$, and one is interested in removing only the immediate junction conditions arising from $R$, a coupling between $T$ and $\mathcal G$ would still be allowed, even if none of the scalars $T$ and $\mathcal G$ is allowed to be coupled to $R$.

\section{Junction conditions with double gravitational layers}\label{sec:double}

Throughout Sec. \ref{sec:juncgen} it was mentioned a few times, namely in the sections associated with the theories $f\left(P\right)$, $f\left(Q\right)$, and $f\left(\mathcal R\right)$, that some of the junction conditions of these theories were imposed with the purpose of avoiding the appearance of double gravitational layers. These double gravitational layers are additional contributions to the stress-energy tensor $T^{\mu\nu}$ that take a distributional form featuring terms dependent on derivatives of the Dirac-$\delta$ distribution function. In this section, we analyze these additional contributions and under which conditions they may arise.

\subsection{Derivatives of the Dirac-$\delta$ distribution function}

Although the use of Gaussian coordinates proves extremely useful in the analysis of junction conditions performed in the previous sections, as they allow one to define the $\delta$ distribution as a function of a single affine parameter $l$, the use of such framework leads to difficulties when dealing with derivatives of $\delta$. To overcome this problem, terms proportional to derivatives of the $\delta$ distribution must be analyzed explicitly through the formalism of distribution functions applied to some test tensor function \cite{senovilla1,Reina:2015gxa}. In this section, we clarify how to proceed to obtain the derivatives of the $\delta$ distribution needed in what follows to analyze double gravitational layers.

Let $T_F$ be the distribution function characterized by the function $F\left(l\right)$. For some suitable test function $Y$ with compact support, the distribution $T_F$ acts on $Y$ according to the following definition:
\begin{equation}
    \left<T_F,Y\right>=\int_\Omega F Y d\Omega.
\end{equation}
Following this definition, the covariant derivative of $T_F$ acting on some tensorial test function $Y^\mu$ of compact support yields
\begin{equation}
    \left<\nabla_\mu T_F, Y^\mu\right>=-\left<T_F,\nabla_\mu Y^\mu\right>,
\end{equation}
where an integration by parts was used and the boundary term vanishes due to the fact that $Y^\mu$ has compact support. Let now $\delta$ be the distribution function associated with the function $\delta\left(l\right)$. Following the result above, the covariant derivative $\nabla_\mu \delta$ thus acts on a test function $Y^\mu$ as
\begin{equation}\label{eq:ddelta1}
    \left<\nabla_\mu \delta, Y^\mu\right>=-\left<\delta,\nabla_\mu Y^\mu\right>=-\left<\delta,\epsilon n_\mu n^\nu \nabla_\nu Y^\mu\right>-\left<\delta,h_\mu^\nu\nabla_\nu Y^\mu\right>,
\end{equation}
where in the last equality we have split the covariant derivative $\nabla_\mu$ into its components orthogonal and tangential to $\Sigma$, i.e., $\epsilon n_\mu n^\nu \nabla_\nu$ and $h_\mu^\nu\nabla_\nu$, respectively, where $h_{\mu\nu}\equiv e_\mu^a e_\nu^b h_{ab}$. The first term on the right-hand side of Eq. \eqref{eq:ddelta1} can be manipulated into the form $\left<\nabla_\mu\left(\epsilon n_\mu n^\nu \delta\right),Y^\mu\right>$, and thus it can be thought of as the application of some tensorial distribution function $\Delta_\mu\equiv \nabla_\nu\left(\epsilon n_\mu n^\nu \delta\right)$ acting on the test function $Y^\mu$ as
\begin{equation}\label{eq:def_Delta1}
    \left<\Delta_\mu, Y^\mu\right>=-\int_\Omega \epsilon\delta\left(l\right)n_\mu n^\nu \nabla_\nu Y^\mu d\Omega=-\int_\Sigma\epsilon n_\mu n^\nu\nabla_\nu Y^\mu d\Sigma.
\end{equation}
On the other hand, the covariant derivative on the second term on the right-hand side of Eq. \eqref{eq:ddelta1} can be dealt with via the application of the Gauss' theorem, from which one obtains $\left<\delta,h_\mu^\nu\nabla_\nu Y^\mu\right>=\left<\delta,\epsilon K^\nu_\nu n_\mu Y^\mu\right>$, where $K_{\mu\nu}^\pm\equiv e_\mu^a e_\nu^b K_{ab}^\pm$. Note that ${K_\nu^\nu}^\pm$ depends on the side of $\Sigma$ where it is computed. Combining this result together with Eq. \eqref{eq:def_Delta1}, one thus obtains
\begin{equation}
    \nabla_\mu\delta\left(l\right) = \nabla_\nu\left(\epsilon n_\mu n^\nu \delta\left(l\right)\right)-\epsilon\left\{K^\nu_\nu\right\}n_\mu\delta\left(l\right)\equiv \Delta_\mu -\epsilon\left\{K^\nu_\nu\right\}n_\mu\delta\left(l\right).
\end{equation}

Consider now the situation for which one is interested in the covariant derivative of the $\delta$ distribution multiplied by some tensor quantity $X_\mu$. Proceeding in a similar fashion and acting on some tensor function of compact support $Y^{\mu\nu}$, one obtains
\begin{equation}\label{eq:ddelta2}
    \left<\nabla_\mu\left(X_\nu \delta\right),Y^{\mu\nu}\right>=-\left<\delta,\epsilon X_\nu n_\mu n^\alpha\nabla_\alpha Y^{\mu\nu}\right>-\left<\delta,X_\nu h_\mu^\alpha\nabla_\alpha Y^{\mu\nu} \right>.
\end{equation}
Similarly as before, the first term on the right-hand side of Eq.\eqref{eq:ddelta2} can be manipulated into the form $\left<\nabla_\alpha\left(\epsilon X_\nu n_\mu n^\alpha\delta\right),Y^{\mu\nu}\right>$, and thus it can be interpreted as the application of some distribution function $\Delta_{\mu\nu}\equiv \nabla_\alpha\left(\epsilon X_\nu n_\mu n^\alpha\delta\right)$ acting on a test function $Y^{\mu\nu}$ as
\begin{equation}\label{eq:def_Delta2}
    \left<\Delta_{\mu\nu}, Y^{\mu\nu}\right>=-\int_\Omega \epsilon \delta\left(l\right)X_\nu n_\mu n^\alpha \nabla_\alpha Y^{\mu\nu} d\Omega=-\int_\Sigma \epsilon X_\nu n_\mu n^\alpha\nabla_\alpha Y^{\mu\nu} d\Sigma.
\end{equation}
On the other hand, the second term on the right-hand side of Eq.\eqref{eq:ddelta2} can be expanded into $\left<\delta,h_\mu^\alpha\nabla_\alpha\left(X_\nu Y^{\mu\nu}\right)\right>-\left<\delta,h_\mu^\alpha Y^{\mu\nu}\nabla_\alpha X_\nu\right>$. The first of these terms can be simplified via the use of the Gauss' theorem into $\left<\delta,\epsilon X_\nu n_\mu K^\alpha_\alpha Y^{\mu\nu}\right>$. Combining this result with Eq. \eqref{eq:def_Delta2}, one thus obtains
\begin{equation}\label{eq:delta_Delta}
    \nabla_\mu\left(X_\nu \delta\left(l\right)\right)=\nabla_\alpha\left(\epsilon X_\nu n_\mu n^\alpha\delta\left(l\right)\right)-\left(\epsilon X_\nu n_\mu \left\{K^\alpha_\alpha\right\}-h_\mu^\alpha \nabla_\alpha X_\nu\right)\delta\left(l\right)\equiv \Delta_{\mu\nu}-\left(\epsilon X_\nu n_\mu \left\{K^\alpha_\alpha\right\}-h_\mu^\alpha \nabla_\alpha X_\nu\right)\delta\left(l\right).
\end{equation}

In what follows, the result of Eq. \eqref{eq:delta_Delta} is essential in the understanding of how the additional contributions to the stress-energy tensor, associated with double gravitational layers and terms non-tangential to $\Sigma$, arise in situations for which some of the previously obtained junction conditions are discarded.

\subsection{Stress-energy tensor in the presence of a double gravitational layer}

To clarify how derivatives of the $\delta$ distribution arise in the theories considered, take as an example the quantity derivative of an arbitrary quantity $X$ given in Eq. \eqref{eq:def_ddist}, which features a term proportional to $\left[X\right]\delta\left(l\right)$. In case there are no junction conditions requiring that $\left[X\right]=0$ to eliminate this term, the second-order derivatives of $X$ would take the form \cite{senovilla1,Reina:2015gxa}
\begin{equation}
    \nabla_\mu\nabla_\nu X= \nabla_\mu\nabla_\nu X^\pm + \epsilon n_\mu \left[\nabla_\nu X\right]\delta\left(l\right)+\nabla_\mu\left(\epsilon n_\nu\left[X\right]\delta\left(l\right)\right).
\end{equation}
Defining the tensor $X_\mu\equiv \epsilon n_\mu\left[X\right]$, one can use the result of Eq. \eqref{eq:delta_Delta} to compute the term $\nabla_\mu\left(n_\nu\left[X\right]\delta\left(l\right)\right)$. Furthermore, the resulting term $h_\mu^\alpha\nabla_\alpha\left(n_\nu\left[X\right]\right)$ can be expanded into two distinct terms, one containing the extrinsic curvature $K_{\mu\nu}^\pm$ and another containing the derivative of $\left[X\right]$. The resultant expression for the second-order covariant derivative of the quantity $X$ is thus
\begin{equation}
    \nabla_\mu\nabla_\nu X= \nabla_\mu\nabla_\nu X^\pm + \epsilon n_\mu \left[\nabla_\nu X\right]\delta\left(l\right)+\epsilon\Delta^X_{\mu\nu}+\epsilon\delta\left(l\right)\left(K_{\mu\nu}-\epsilon \left\{K^\alpha_\alpha\right\} n_\mu n_\nu + 2 h^\sigma_{(\mu} n_{\nu)} \nabla_\sigma\right)\left[X\right],
\end{equation}
where we have defined $K_{\mu\nu}^\pm=e^a_\mu e^b_\nu K_{ab}^\pm$ and $h_{\mu\nu}^\pm=e^a_\mu e^b_\nu h_{ab}^\pm$, and  $\Delta^X_{\mu\nu}=\nabla_\alpha\left(\epsilon \left[X\right] n_\nu n_\mu n^\alpha\delta\left(l\right)\right)$ is a distribution function that can be written implicitly for any test function $Y^{\mu\nu}$ with compact support as
\begin{equation}
    \left<\Delta_{\mu\nu}^X,Y^{\mu\nu}\right>= - \int_\Sigma \epsilon \left[X\right]n_\mu n_\nu n^\sigma\nabla_\sigma Y^{\mu\nu} d\Sigma.
\end{equation}
Note that taking $\left[X\right]=0$ one recovers the usual regular second-order derivatives of the quantity $X$ that have appeared throughout this manuscript several times in Sec. \ref{sec:juncgen}. The fact that $\left[X\right]\neq 0$ implies that additional contributions to the stress-energy tensor $T^{\mu\nu}$ arise, not only in the form of additional contributions to the stress-energy tensor of the thin-shell $S_{ab}$ defined on $\Sigma$, but also including additional off-shell terms proportional the normal vector $n_\mu$, and the so-called double gravitational layer described by the distribution $\Delta_{\mu\nu}$. When such a situation arises, the stress-energy tensor in the distribution formalism is written in the form
\begin{equation}\label{eq:def_Tab_double}
T_{\mu\nu} =T_{\mu\nu}^+\Theta\left(l\right)+T_{\mu\nu}^-\Theta\left(-l\right)+\delta\left(l\right)\left(S_{\mu\nu}+2S_{(\mu}n_{\nu)}+S n_\mu n_\nu\right)+s_{ab},
\end{equation}
where $S_{\mu\nu}$ is the stress-energy tensor of the thin-shell, $S_\mu$ represents an external momentum flux whose normal component represents the energy flux across $\Sigma$ and spatial components represent the tangential stresses, $S$ represents an external normal tension supported on $\Sigma$, and $s_{ab}$ is the stress-energy tensor of the double gravitational layer. 

\subsection{Appearance of double gravitational layers}

In Sec. \ref{sec:eliminatediff} it was mentioned that differential junction conditions and the contributions to the stress-energy tensor of the thin shell in a given theory $f\left(X\right)$ can be discarded by taking a particular form of the action for which the terms $\nabla_\mu X\nabla_\nu X$ and $\nabla_\mu\nabla_\nu X$ are removed from the field equation, respectively. A particular case of interest arises when the choice of the function $f\left(X\right)$ is such as the term $\nabla_\mu X\nabla_\nu X$ is removed from the field equation, thus discarding the differential junction conditions associated with the scalar $X$, but the term $\nabla_\mu\nabla_\nu X$ remains present. In such a case, the absence of a differential junction condition leads to the appearance of a $\delta\left(l\right)$ term in $\nabla_\mu X$ which, consequently, leads to the appearance of the additional contributions to the stress-energy tensor mentioned previously. 

Another situation under which double gravitational layers may arise independently of the form chosen for the function $f\left(X\right)$ is through differential terms in the tensors $H_{\mu\nu}^{(X)}$ proportional to derivatives of geometrical or matter quantities without derivatives of the function $f_X$. Such a situation appears e.g. in the $f\left(Q\right)$ theory, where the tensor $H_{\mu\nu}^{(Q)}$ features a term proportional to derivatives of ${R^\sigma_{\ (\mu\nu)}}^\rho$; and also in the $f\left(\mathcal R\right)$ theory, where the tensor $H_{\mu\nu}^{(\mathcal R)}$ features terms proportional to derivatives of $T^{\mu\nu}$. We note however that contributions to the double gravitational layer arising this way, i.e., from terms that do not consist of derivatives of the function $f_X$, do not induce additional coupling contributions when the scalar $X$ is coupled to another scalar $Y$ (similarly to what happens for the immediate junction conditions in comparison with the differential junction conditions in Sec. \ref{sec:juncgen}). 

Let us now investigate how the different scalars $X_i$ contribute to the rise of double gravitational layers. Later, in Sec. \ref{sec:Cdouble}, we analyze how the couplings between two scalars $X_i$ and $X_j$ affect these results.

\subsubsection{Contribution of $f\left(R\right)$}

We note again that the junction conditions with double gravitational layers for the $f\left(R\right)$ theory have already been analyzed in detail in Ref. \cite{senovilla1}, but we chose to include them in this manuscript as well to guarantee its self-consistency. For the $f\left(R\right)$ theory, the tensor $H_{\mu\nu}^{(R)}$ depends on differential terms proportional to derivatives of the function $f_R$, see Eq. \eqref{eq:def_H_R}. These terms can be expanded in terms of derivatives of $R$ as given in Eq. \eqref{eq:R_ddf}. From this equation, it can be seen that if $f_{RRR}=0$, then the derivatives of the form $\nabla_\mu R\nabla_\nu R$, responsible for the appearance of the junction condition $\left[R\right]=0$, are removed from the field equations. Consequently, the terms proportional to $\nabla_\mu\nabla_\nu R$ are altered and feature contributions to the double gravitational layer:
\begin{equation}\label{eq:R_DL}
        \nabla_\mu\nabla_\nu R= \nabla_\mu\nabla_\nu R^\pm + \epsilon n_\mu \left[\nabla_\nu R\right]\delta\left(l\right)+\epsilon\Delta^R_{\mu\nu}+\epsilon\delta\left(l\right)\left(\{K_{\mu\nu}\}-\epsilon \{K^\alpha_\alpha\} n_\mu n_\nu + 2h^\sigma_{(\mu} n_{\nu)} \nabla_\sigma\right)\left[R\right],
\end{equation}
where the distribution function $\Delta^R_{\mu\nu}$ is defined as
\begin{equation}
\Delta^R_{\mu\nu}=\nabla_\alpha\left(\epsilon \left[R\right] n_\nu n_\mu n^\alpha\delta\left(l\right)\right).
\end{equation}
Taking the field equations in Eq.\eqref{eq:field} with $H_{\mu\nu}$ given by Eq.\eqref{eq:def_H_R}, with the quantities in the distribution formalism given in Eqs.\eqref{eq:dist_Rab}, \eqref{eq:dist_R} and \eqref{eq:R_DL}, this time subjected only to the immediate junction condition in Eq.\eqref{eq:R_jc_1}, one obtains the following additional contributions to the stress energy tensor in comparison to the case for which $\left[R\right]=0$:
\begin{equation}\label{eq:doubleR1}
    8\pi S_{ab}=-\epsilon f_{RR} \left\{K_{ab}\right\}\left[R\right],
\end{equation}
\begin{equation}
    8\pi S_\mu = -2\epsilon f_{RR}h_\mu^\nu\nabla_\nu\left[R\right],
\end{equation}
\begin{equation}
    8\pi S=f_{RR} \left\{K\right\} \left[R\right],
\end{equation}
\begin{equation}\label{eq:doubleR2}
    8\pi s_{ab}= \epsilon f_{RR} \left(h_{ab}\Delta^R-e^\mu_a e^\nu_b\Delta_{\mu\nu}^R\right),
\end{equation}
where $S_{ab}$ and $s_{ab}$ are obtained by taking the projection of the field equations into the hypersurface $\Sigma$ with $e_a^\mu s_b^\nu$, $\Delta^R\equiv g^{\mu\nu}\Delta^R_{\mu\nu}$.

\subsubsection{Contribution of $f\left(P\right)$}

Regarding the $f\left(P\right)$ theory, the tensor $H^{(P)}_{\mu\nu}$ depends on derivatives of both the function $f_P$ and the Ricci tensor $R_{\mu\nu}$, the latter under sever different contractions, see Eq.\eqref{eq:def_H_P}. The derivtives of the function $f_P$ can be expanded in terms of derivatives of $P$ via Eq. \eqref{eq:P_ddf}, from which one verifies that, if $f_{PPP}=0$, then the terms of the form $\nabla_\mu P \nabla_\nu P$, which previously were responsible for the junction condition $\left[P\right]=0$, are effectively removed from the field equations. As a result, the second-order covariant derivatives $\nabla_\mu\nabla\nu P$ now feature additional terms potentially leading to double gravitational layers, namely
\begin{eqnarray}
        \nabla_\mu\nabla_\nu P= \nabla_\mu\nabla_\nu P^\pm + \epsilon n_\mu \left[\nabla_\nu P\right]\delta\left(l\right)+ \epsilon\Delta^P_{\mu\nu}+ \epsilon\delta\left(l\right)\left(\{K_{\mu\nu}\}-\epsilon \{K^\alpha_\alpha\} n_\mu n_\nu + 2h^\sigma_{(\mu} n_{\nu)} \nabla_\sigma\right)\left[P\right],
\end{eqnarray}
where we have defined the distribution function $\Delta_{\mu\nu}^P$ as 
\begin{equation}
\Delta^P_{\mu\nu}=\nabla_\alpha\left(\epsilon \left[P\right] n_\nu n_\mu n^\alpha\delta\left(l\right)\right).
\end{equation}
Similarly, we have previously enforced the junction condition $\left[R_{\mu\nu}\right]=0$ to guarantee that any additional terms contributing to the double gravitational layer were removed from the second-order covariant derivatives of $R_{\mu\nu}$. In this section, we thus discard this non-essential junction condition, and allow for the first derivative of $R_{\mu\nu}$ to depend on $\left[R_\mu\nu\right]$. The relaxation of this condition implies that the terms in the field equations proportional to products of, e.g., $\nabla_\alpha R_{\mu\nu}\nabla_\beta f_P$ can now feature singular products of the form $\left[R_{\mu\nu}\right]\left[P\right]\delta^2\left(l\right)$. To preserve the regularity of the field equations, one must thus impose either that $\left[R_\mu\nu\right]=0$ or that $\left[P\right]=0$. Given that the scalar $P$ is defined as $P=R_{\mu\nu}R^{\mu\nu}$, the condition $\left[R_{\mu\nu}\right]=0$ immediately forces that $\left[P\right]=0$ via the property of the jump given in Eq. \eqref{eq:def_prop}, which results in an immediate absence of any terms potentially leading to a double gravitational layer. However, the inverse property is not true, i.e., the imposition of the condition $\left[P\right]=0$ does not imply that $\left[R_{\mu\nu}\right]$, e.g. if it happens that $\left\{R_{\mu\nu}\right\}=0$. Thus, to allow for the survival of some of these potentially interesting terms, we consider in what follows that $\left[P\right]=0$ and $\left[R_{\mu\nu}\right]\neq 0$. In this case, the double gravitational layers may still arise but induced solely by the differential terms of $R_{\mu\nu}$.

Consider thus the following most general forms of the first and second-order covariant derivatives of $R_{\mu\nu}$:
\begin{equation}
    \nabla_\sigma R_{\mu\nu}=\left(\nabla_\sigma R_{\mu\nu}\right)^\pm+\epsilon n_\sigma\left[R_{\mu\nu}\right]\delta\left(l\right),
\end{equation}
\begin{equation}\label{eq:Rab_double}
    \nabla_\sigma\nabla_\rho R_{\mu\nu}=\left(\nabla_\sigma\nabla_\rho R_{\mu\nu}\right)^\pm+\epsilon n_\sigma\left[\nabla_\rho R_{\mu\nu}\right]\delta\left(l\right)+\epsilon \Delta^{\rm Ric}_{\sigma\rho\mu\nu}+\epsilon\delta\left(l\right)\left(\left\{K_{\sigma\rho}\right\}-\epsilon\left\{K^\alpha_\alpha\right\}n_\sigma n_\rho+2h^\alpha_{(\sigma}n_{\rho)}\nabla_\alpha\right)\left[R_{\mu\nu}\right],
\end{equation}
where the distribution function $\Delta^{\rm Ric}_{\sigma\rho\mu\nu}$ is defined as
\begin{equation}
\Delta^{\rm Ric}_{\sigma\rho\mu\nu}=\nabla_\alpha\left(\epsilon \left[ R_{\mu\nu}\right]n_\sigma n_\rho n^\alpha \delta\left(l\right)\right).
\end{equation}
Taking thus the field equations given in Eq.\eqref{eq:field} with the distribution functions in Eq. \eqref{eq:dist_Rab} and \eqref{eq:dist_P}, along with the junction conditions in Eqs. \eqref{eq:P_jc_1} and \eqref{eq:P_jc_2}, but not Eq. \eqref{eq:P_jc_3}, and taking the result for the second-order covariant derivative of $R_{\mu\nu}$ obtained through Eq. \eqref{eq:Rab_double}, and projecting the result into $\Sigma$ through the use of $e^\mu_a e^\nu_b$, one obtains the following additional contributions in comparison to the case without double layers:
\begin{equation}
8\pi S_{ab}=\epsilon f_P\left\{h_{ab}\left\{K_{\mu\nu}\right\}\left[R^{\mu\nu}\right]+e_a^\mu e_b^\nu\left[\left\{K^\gamma_\gamma\right\}\left[R_{\mu\nu}\right]-2\left\{K_{(\mu\gamma}\right\}\left[R^\gamma_{\nu)}\right]-2h^\gamma_{(\mu}n_\alpha\nabla_\gamma\left[R^\alpha_{\nu)}\right]\right]\right\},
\end{equation}
\begin{equation}
    8\pi S_\mu=2 f_P\left(\left\{K^\gamma_\gamma\right\}n_\alpha-\epsilon h^\gamma_\alpha\nabla_\gamma\right)\left[R^\alpha_\mu\right],
\end{equation}
\begin{equation}
    8\pi S=0,
\end{equation}
\begin{equation}
8\pi s_{ab}=\epsilon f_P\left\{h_{ab}\left({\Delta_{\mu\nu}}^{\mu\nu}\right)^{\rm Ric}+e_a^\mu e_b^\nu\left[\left({\Delta^\gamma}_{\gamma\mu\nu}\right)^{\rm Ric}-2\left({\Delta_{(\mu\gamma\nu)}}^\gamma\right)^{\rm Ric}\right]\right\}.
\end{equation}

Note that the contributions obtained to the double gravitational layer arise from differential terms that do not contain higher-order derivatives of the function $f_P$. This implies that these terms are not prone to couplings between $P$ and other scalars, as discussed in Sec. \ref{sec:Cdouble}. Furthermore, note also that if the function $f\left(P\right)$ is linear in $P$, which consequently implies that $f_Q$ is a constant and thus $\nabla_\mu f_Q=0$, the terms proportional to $\nabla_\sigma f_P\nabla_\rho R_{\mu\nu}$, which are responsible for the junction condition $\left[P\right]=0$, vanish from the field equation, and thus one should expect additional contributions to the stress-energy tensor of the thin-shell and to the double gravitational layer in such a case.

\subsubsection{Contribution of $f\left(Q\right)$}

For the $f\left(Q\right)$ theory, the tensor $H_{\mu\nu}^{(Q)}$ depends on differential terms proportional to derivatives of the function $f_Q$, and derivatives of the quantity ${R^\sigma_{\ (\mu\nu)}}^\rho$, see Eq. \eqref{eq:def_H_R}. The first of these terms can be expanded in terms of derivatives of $Q$ as given in Eq. \eqref{eq:Q_ddf}, from which one verifies that if $f_{QQQ}=0$, then the derivatives of the form $\nabla_\mu Q\nabla_\nu Q$, which lead to the junction condition $\left[Q\right]=0$, are removed from the field equations. Consequently, the terms proportional to $\nabla_\mu\nabla_\nu Q$ are altered and feature potential contributions to the double gravitational layer:
\begin{eqnarray}
        \nabla_\mu\nabla_\nu Q= \nabla_\mu\nabla_\nu Q^\pm + \epsilon n_\mu \left[\nabla_\nu Q\right]\delta\left(l\right)+ \epsilon\Delta^Q_{\mu\nu}+ \epsilon\delta\left(l\right)\left(\{K_{\mu\nu}\}-\epsilon \{K^\alpha_\alpha\} n_\mu n_\nu + 2h^\sigma_{(\mu} n_{\nu)} \nabla_\sigma\right)\left[Q\right],
\end{eqnarray}
where the distribution function $\Delta^Q_{\mu\nu}$ is defined as
\begin{equation}
\Delta^Q_{\mu\nu}=\nabla_\alpha\left(\epsilon \left[Q\right] n_\nu n_\mu n^\alpha\delta\left(l\right)\right).
\end{equation}
On the other hand, the first derivative of ${R^\sigma_{\ (\mu\nu)}}^\rho$ is now allowed to depend on $\left[{R^\sigma_{\ (\mu\nu)}}^\rho\right]\delta\left(l\right)$, unlike it happened in Sec. \ref{sec:juncgen} where such terms were forced to vanish to avoid the appearance of the double gravitational layer. This implies that the crossed terms $\nabla_\sigma {R^\sigma_{\ (\mu\nu)}}^\rho\nabla_\rho f_Q$, if $\left[Q\right]$ is allowed to be non-zero, would feature terms proportional to $\left[{R^\sigma_{\ (\mu\nu)}}^\rho\right]\left[Q\right]\delta^2\left(l\right)$, which are singular in the distribution formalism. To avoid the presence of these terms, one can either force $\left[Q\right]=0$ or $\left[{R^\sigma_{\ (\mu\nu)}}^\rho\right]=0$. Given that the scalar $Q$ is defined as $Q=R_{\mu\nu\sigma\rho}R^{\mu\nu\sigma\rho}$, if one considers the option $\left[{R^\sigma_{\ (\mu\nu)}}^\rho\right]=0$, this implies via the property of the jump given in Eq. \eqref{eq:def_prop} that $\left[Q\right]=0$, thus eliminating every term that could potentially lead to a double gravitational layer (similarly to what happens for the $f\left(P\right)$ theory). However, the inverse is not true, i.e., if one considers the option that $\left[Q\right]=0$, this does not imply that $\left[{R^\sigma_{\ (\mu\nu)}}^\rho\right]=0$ as, by the same property (see Eq. \eqref{eq:def_prop}), it is possible that $\left\{{R^\sigma_{\ (\mu\nu)}}^\rho\right\}=0$ and $\left[{R^\sigma_{\ (\mu\nu)}}^\rho\right]\neq 0$. In such a case, the double gravitational layer can still arise but induced by the differential terms in ${R^\sigma_{\ (\mu\nu)}}^\rho$, instead of the differential terms in $Q$. 

Let us thus consider the following forms for the first and second-order covariant derivatives of ${R^\sigma_{\ (\mu\nu)}}^\rho$:
\begin{equation}
    \nabla_\sigma {R^\sigma_{\ (\mu\nu)}}^\rho=\left(\nabla_\sigma {R^\sigma_{\ (\mu\nu)}}^\rho\right)^\pm + \epsilon n_\sigma \left[{R^\sigma_{\ (\mu\nu)}}^\rho\right]\delta\left(l\right),
\end{equation}
\begin{equation}\label{eq:Rabcd_double}
        \nabla_\sigma\nabla_\rho {R^\sigma_{\ (\mu\nu)}}^\rho= \left(\nabla_\sigma\nabla_\rho {R^\sigma_{\ (\mu\nu)}}^\rho\right)^\pm + \epsilon n_\sigma \left[\nabla_\rho {R^\sigma_{\ (\mu\nu)}}^\rho\right]\delta\left(l\right)+\epsilon \Delta^{\rm Rie}_{\mu\nu}+\epsilon\delta\left(l\right)\left(\{K_{\sigma\rho}\}-\epsilon \{K^\alpha_\alpha\} n_\sigma n_\rho + 2h^\alpha_{(\sigma} n_{\rho)} \nabla_\alpha\right)\left[{R^\sigma_{\ (\mu\nu)}}^\rho\right], 
\end{equation}
where the distribution function $\Delta^{\rm Rie}_{\mu\nu}$ is defined as
\begin{equation}
    \Delta^{\rm Rie}_{\mu\nu}=    \nabla_\alpha\left(\epsilon \left[{R^\sigma_{\ (\mu\nu)}}^\rho\right] n_\sigma n_\rho n^\alpha\delta\left(l\right)\right).
\end{equation}
We note that, in contrast with what happens for the $f\left(R\right)$ theory, in this case the additional contributions to the stress energy tensor are all given at the shell, i.e., there are no external fluxes and tensions on $\Sigma$. This happens due to the fact that the normal vectors $n_\sigma$ and $n_\rho$ in Eq. \eqref{eq:Rabcd_double} are contracted with ${R^\sigma_{\ (\mu\nu)}}^\rho$, whereas before they were free vectors. Taking the field equations in Eq.\eqref{eq:field} with the distribution functions in Eqs.\eqref{eq:dist_Rabcd} and \eqref{eq:dist_Q}, along with the junction conditions in Eqs.\eqref{eq:Q_jc_1} and \eqref{eq:Q_jc_2} but not \eqref{eq:Q_jc_3}, and considering now the result obtained for the second-order covariant derivative of ${R^\sigma_{\ (\mu\nu)}}^\rho$ in Eq. \eqref{eq:Rabcd_double} and projecting the result into $\sigma$ with , we obtain the additional contributions
\begin{equation}
    8\pi S_{ab}=-8\epsilon f_{QQ}e^\mu_a e^\nu_b n_{(\sigma}\left\{\nabla_{\rho)}Q\right\}\left[{R^\sigma_{\ (\mu\nu)}}^\rho\right]-4\epsilon f_Qe^\mu_a e^\nu_b\left(\left\{K_{\sigma\rho}\right\}-\epsilon \left\{K\right\} n_\sigma n_\rho+2 n_{(\sigma} h^\alpha_ {\rho)}\nabla_\alpha\right)\left[{R^\sigma_{\ (\mu\nu)}}^\rho\right],
\end{equation}
\begin{equation}
    8\pi s_{ab}=-4\epsilon f_Q e^\mu_a e^\nu_b\Delta_{\mu\nu}^{\rm Rie}.
\end{equation}
We note that, since these contributions to the double gravitational layer arise from differential terms that do not contain higher-order derivatives of the function $f_Q$, they are not prone to couplings between $Q$ and other scalars, as discussed in Sec. \ref{sec:Cdouble}. Furthermore, note that if the function $f\left(Q\right)$ is linear in $Q$, which implies that $f_Q$ is a constant and thus $\nabla_\mu f_Q=0$, the terms proportional to $\nabla_\sigma f_Q\nabla_\rho {R^\sigma_{\ (\mu\nu)}}^\rho$, which force $\left[Q\right]=0$, are be absent from the field equation, and thus one could have additional contributions to the stress-energy tensor of the thin-shell and to the double gravitational layer.

\subsubsection{Contribution of $f\left(\mathcal G\right)$}

Consider now the $f\left(\mathcal G\right)$ theory. The tensor $H_{\mu\nu}^{(\mathcal G)}$ depends on differential terms proportional to derivatives of the function $f_\mathcal G$, see Eq. \eqref{eq:def_H_G}. These terms can be expanded in terms of derivatives of $\mathcal G$ as given in Eq. \eqref{eq:G_ddf}. One verifies that if $f_{\mathcal G\mathcal G\mathcal G}=0$, then the differential terms $\nabla_\mu \mathcal G\nabla_\nu \mathcal G$, which cause the appearance of the junction condition $\left[\mathcal G\right]=0$, are eliminated from the field equations. Consequently, the terms proportional to $\nabla_\mu\nabla_\nu \mathcal G$ are include contributions to the double gravitational layer:
\begin{eqnarray}\label{eq:G_double}
        \nabla_\mu\nabla_\nu \mathcal G= \nabla_\mu\nabla_\nu \mathcal G^\pm + \epsilon n_\mu \left[\nabla_\nu \mathcal G\right]\delta\left(l\right)+\epsilon \Delta^{\mathcal G}_{\mu\nu}+\epsilon\delta\left(l\right)\left(\{K_{\mu\nu}\}-\epsilon \{K^\alpha_\alpha\} n_\mu n_\nu + 2h^\sigma_{(\mu} n_{\nu)} \nabla_\sigma\right)\left[\mathcal G\right],
\end{eqnarray}
where the distribution function $\Delta^{\mathcal G}_{\mu\nu}$ is defined as
\begin{equation}
\Delta^{\mathcal G}_{\mu\nu}=    \nabla_\alpha\left(\epsilon \left[\mathcal G\right] n_\mu n_\nu n^\alpha\delta\left(l\right)\right)
\end{equation}
Similarly to what happens in the $f\left(Q\right)$ theory, since in the field equations the indices of the derivatives $\nabla^\sigma \nabla^\rho f_\mathcal G$ are not free indices but, instead, are contracted with the geometrical quantities that are factorized before, this implies that every additional contribution to the stress-energy tensor is defined only at the hypersurface $\Sigma$, because no free normal vectors are present in the expression. Taking the field equations in Eq.\eqref{eq:field} with a $H_{\mu\nu}$ tensor given by Eq.\eqref{eq:def_H_G}, with the quantities in the distribution formalism given in Eq.\eqref{eq:dist_Rabcd}, \eqref{eq:dist_Rab}, and \eqref{eq:dist_R}, but now considering the second-order derivatives of $\mathcal G$ as given by Eq. \eqref{eq:dist_ddG}, considering solely the junction condition in Eq.\eqref{eq:G_jc_1}, and projecting the result into $\Sigma$ using $e^\mu_a e^\nu_b$, one obtains the additional contributions to the stress energy tensor
\begin{equation}
8\pi S_{ab}=4\epsilon e^\mu_a e^\nu_b f_{\mathcal G\mathcal G}\left(K^{\sigma\rho}-\epsilon K n^\sigma n^\rho+h^{\sigma\alpha}n^\rho \nabla_\alpha\right)\left[\mathcal G\right]\left( \left\{R_{\mu\sigma\nu\rho}\right\}+2g_{\sigma[\nu}\left\{R_{\rho]\mu}\right\}+2g_{\mu[\sigma}\left\{G_{\nu]\rho}\right\}\right),
\end{equation}
\begin{equation}
    8\pi s_{ab}=4\epsilon e^\mu_a e^\nu_b f_{\mathcal G\mathcal G}\Delta^{\sigma\rho}_{\mathcal G}\left( \left\{R_{\mu\sigma\nu\rho}\right\}+2g_{\sigma[\nu}\left\{R_{\rho]\mu}\right\}+2g_{\mu[\sigma}\left\{G_{\nu]\rho}\right\}\right).
\end{equation}

\subsubsection{Contribution of $f\left(T\right)$}

For the $f\left(T\right)$ theory, and given that the tensor $H_{\mu\nu}^{(T)}$ does not feature any differential terms proportional to derivatives of the function $f_T$, no double gravitational layers can arise from the theory $f\left(T\right)$ only. Nevertheless, and given that the junction conditions in this theory do not require that $\left[T\right]=0$, it is still possible that double gravitational layers depending on $\left[T\right]$ may arise if the scalar $T$ is coupled to some other scalar. Such an example is given in what follows in Sec. \ref{sec:Cdouble}.

\subsubsection{Contribution of $f\left(\mathcal T\right)$}

Similarly to what happens for the $f\left(T\right)$ theory, given that the $f\left(\mathcal T\right)$ features a tensor $H_{\mu\nu}^{(\mathcal T)}$ with no differential terms proportional to derivatives of the function $f_\mathcal{T}$, no double gravitational layers can arise from the $f\left(\mathcal T\right)$ theory. There is, however, an important distinction between what happens for the $f\left(T\right)$ and the $f\left(\mathcal T\right)$ theories. Since the scalar $\mathcal T$ is defined as $\mathcal T=T_{\mu\nu}T^{\mu\nu}$, the regularity of $\mathcal T$ requires that no singular terms can appear in $T_{\mu\nu}$ (see Sec. \ref{sec:juncgen}). Since the additional contributions to the stress-energy tensor and the double gravitational layer correspond to singular terms of $T_{\mu\nu}$, see Eq. \eqref{eq:def_Tab_double}, such a regularity immediately implies that every singular component of the stress-energy tensor must vanish, i.e.,
\begin{equation}
    S_{ab}=0,\quad S_a=0,\quad S=0,\quad s_{ab}=0.
\end{equation}
Thus, unlike in the $f\left(T\right)$ theory for which double gravitational layers with terms proportional to $\left[T\right]$ can be induced by couplings with other scalars, in the $f\left(\mathcal T\right)$ theory not only such couplings are not possible but also the regularity of the scalar $\mathcal T$ forces all components of the double gravitational layer to vanish, even if couplings with other scalars are present. 

\subsubsection{Contribution of $f\left(\mathcal R\right)$}

For the $f\left(\mathcal R\right)$ theory, even though the tensor $H_{\mu\nu}^{(\mathcal R)}$ features differential terms proportional to derivatives of $f_\mathcal{R}$, double gravitational layers are not allowed. To understand why this happens, recall that the scalar $\mathcal R$ is defined as a product between the Ricci scalar and the stress-energy tensor, i.e., $\mathcal R=R_{\mu\nu}T^{\mu\nu}$, which implies that the regularity of this term requires that no singular terms can appear in either $R_{\mu\nu}$ and $T^{\mu\nu}$ (see Sec. \ref{sec:juncgen}). Similarly to what happens in $f\left(\mathcal T\right)$, since the additional contributions to the stress-energy tensor induced by the double gravitational layer correspond to singular terms of $T^{\mu\nu}$, the regularity of $\mathcal R$ immediately sets every additional singular component of the stress-energy tensor to zero, i.e., one obtains
\begin{equation}
    S_{ab}=0,\quad S_a=0,\quad S=0,\quad s_{ab}=0.
\end{equation}
Such a restriction also implies that the couplings between $\mathcal R$ and other scalars do not induce additional contributions to the stress-energy tensor nor double gravitational layers, similarly to what is discussed for the $f\left(\mathcal T\right)$ theory.

\subsection{Couplings of double gravitational layers}\label{sec:Cdouble}

As mentioned in the previous sections, only the additional contributions to the stress-energy tensor arising from differential terms proportional to derivatives of the functions $f_{X_i}$ may induce additional coupling contributions when the scalar $X_i$ is coupled to another scalar $X_j$. Thus, in this section, we focus on the contributions arising through that procedure.

If the scalar $X$ is coupled to a scalar $Y$ through some function $f\left(X,Y\right)$, which causes terms of the form $\nabla_\mu\nabla_\nu Y$, $\nabla_\mu X\nabla_\nu Y$, and $\nabla_\mu Y\nabla_\nu Y$ to appear in the field equations (see Eq. \eqref{eq:Cddf}), the explicit form of the function $f\left(X,Y\right)$ is crucial to the appearance of double gravitational layers. If the form of the function $f\left(X,Y\right)$ is such that the terms proportional to $\nabla_\mu X\nabla_\nu X$ vanish, e.g. by setting $f_{XXX}=0$, but the terms proportional to $\nabla_\mu X\nabla_\nu Y$ and $\nabla_\mu Y\nabla_\nu Y$ are still present, e.g. by setting $f_{XXY}\neq 0$ and $f_{XYY}\neq 0$, the first of these terms guarantees that both $\left[X\right]=0$ and $\left[Y\right]=0$, thus preventing the double gravitational layer to appear. If instead one considers that $f_{XXY}=0$ while keeping $f_{XYY}\neq 0$, thus eliminating the term proportional to $\nabla_\mu X\nabla_\nu Y$, the double gravitational layer does appear and the additional contributions to the stress-energy tensor take the same forms as the ones for a function $f\left(X\right)$ only. Finally, if the function $f\left(X,Y\right)$ is such that all terms proportional to $\nabla_\mu\nabla_\nu Y$, $\nabla_\mu X\nabla_\nu Y$, and $\nabla_\mu Y\nabla_\nu Y$ vanish, e.g. by setting $f_{XXX}=f_{XXY}=f_{XYY}=0$, then this implies that not only the double gravitational layer is present, but also that it contains contributions both from the $\left[X\right]$ scalar arising via the term $\nabla_\mu\nabla_\nu X$ proportional to $f_{XX}$, and from the $\left[Y\right]$ scalar via the term $\nabla_\mu\nabla_\nu Y$ proportional to $f_{XY}$. In such a situation, the contribution induced by the scalar $Y$ is of the same form as the contribution induced by the scalar $X$ but with a different factor $f_{XY}$ instead of $f_{XX}$.

Summarizing the statements traced in the previous paragraph, for any theory of gravity for which the action depends on more than one scalar $X_i$, if the scalar $X_i$ is such that the tensor $H_{\mu\nu}^{(X)}$ features terms proportional to second-order covariant derivatives of the function $f_X$ and the scalar $X_i$ is coupled to another scalar $X_j$, one verifies that:
\begin{enumerate}
    \item If $f_{X_iX_iX_i}=0$ but $f_{X_iX_iX_j}\neq 0$ and $f_{X_iX_jX_j}\neq 0$, no double gravitational layers arise, even if they would if the function $f\left(X_i\right)$ depended on $X_i$ only;
    \item If $f_{X_iX_iX_i}=f_{X_iX_iX_j}=0$ and $f_{X_iX_jX_j}\neq 0$, the double gravitational layer arises and the corresponding contributions to the stress-energy tensor are of the same form as if the function $f\left(X_i\right)$ depended on $X_i$ only;
    \item If $f_{X_iX_iX_i}=f_{X_iX_iX_j}=f_{X_iX_jX_j}=0$, the double gravitational layer arises and the corresponding contributions to the stress-energy tensor are the ones in the item above plus additional coupling contributions for which the scalar $\left[X_i\right]$ is replaced by the scalar $\left[X_j\right]$ and the function $f_{X_iX_i}$ is replaced by the function $f_{X_iX_j}$.
\end{enumerate}

\subsubsection{Example of couplings: $f\left(R,T\right)$ gravity}

To illustrate how the couplings described above work, let us consider the example of $f\left(R,T\right)$ gravity. In particular, let us assume that the function satisfies the conditions $f_{RRR}=f_{RRT}=f_{RTT}=0$, such that couplings between the scalars $R$ and $T$ induce additional contributions to the stress-energy tensor $T^{\mu\nu}$. The most general form of the function $f\left(R,T\right)$ satisfying the conditions above is
\begin{equation}
    f\left(R,T\right)=\alpha R^2+\beta RT + g\left(T\right),
\end{equation}
where $\alpha$ and $\beta$ are arbitrary constants and $g\left(T\right)$ is a function of $T$ only. For such a theory, one verifies that the only terms that survive in the second-order derivatives of $f_R$, see Eq. \eqref{eq:df_part_RT}, are $f_{RR}\nabla_\mu\nabla\nu R$ and $f_{RT}\nabla_\mu\nabla_\nu T$. The first of these terms features the contributions to the stress-energy tensor given in Eqs. \eqref{eq:doubleR1} to \eqref{eq:doubleR2}. The second term is completely analogous to the first, but the dependencies in $\left[R\right]$ and $f_{RR}$ are replaced by dependencies in $\left[T\right]$ and $f_{RT}$, according to item number 3 above. Thus, in such a case, the contributions to the stress-energy tensor of the thin-shell become
\begin{equation}
    8\pi S_{ab}=-\epsilon \left\{K_{ab}\right\}\left(f_{RR} \left[R\right]+f_{RT} \left[T\right]\right),
\end{equation}
\begin{equation}
    8\pi S_\mu = -2\epsilon h_\nu^\mu \left(f_{RR}\nabla_\mu\left[R\right]+f_{RT}\nabla_\mu\left[T\right]\right),
\end{equation}
\begin{equation}
    8\pi S=\left\{K\right\}\left(f_{RR}  \left[R\right]+f_{RT}  \left[T\right]\right),
\end{equation}
\begin{equation}
    8\pi s_{ab}=\epsilon \left[f_{RR}\left(h_{ab}\Delta^R-\Delta_{ab}^R\right)+f_{RT}\left(h_{ab}\Delta^T-\Delta_{ab}^T\right)\right],
\end{equation}
where we have defined the distribution $\Delta^T_{\mu\nu}=\nabla_\alpha\left(\epsilon \left[T\right] n_\mu n_\nu n^\alpha\delta\left(l\right)\right)$. These results correspond to the ones obtained previously in Ref. \cite{rosafrt}. Note that if we had chosen a function $f\left(R,T\right)$ such that $f_{RTT}\neq 0$, the terms proportional to $\nabla_\mu T\nabla_\nu T$ in $\nabla_\mu\nabla_\nu f_R$ would force $\left[T\right]=0$, thus recovering the contributions to the stress energy tensor of the function $f\left(R\right)$ alone, in accordance with item number 2 above, and also that if we had chosen a function such that $f_{RRT}\neq 0$, the terms proportional to $\nabla_\mu R\nabla_\nu T$ in $\nabla_\mu\nabla_\nu f_R$ would force $\left[R\right]=0$, thus eliminating every contribution to the stress-energy tensor of the thin-shell, in accordance with item number 1 above.

\section{Scalar-tensor representations}\label{sec:strep}

For theories of gravity for which the action $S$ depends on a function of several scalars $f\left(X_1,...,X_n\right)$, it is sometimes convenient to consider a dynamically equivalent scalar-tensor representation of the theory, in which the arbitrary dependence of the action in the scalars $X_i$ is replaced by a set of scalar fields $\phi_i$. The advantage of such a transformation is a potential reduction of the order of the field equations, at the cost of additional equations of motion for the scalar fields. Thus, to finalize the analysis of the junction conditions, let us briefly mention what is the impact of such a transformation for the junction conditions of the theory. 

To deduce a scalar-tensor representation of the theory described in Eq. \eqref{eq:action}, we introduce a set of auxiliary fields $\alpha_i$ in such a way that the action $S$ can be rewritten in the form
\begin{equation}\label{eq:actionST}
S=\frac{1}{2\kappa^2}\int_\Omega\sqrt{-g}\left[R+\sum_{i=1}^n \frac{\partial f}{\partial \alpha_i}\left(X_i-\alpha_i\right)+f\left(\alpha_1,...,\alpha_n\right)+2\kappa^2\mathcal L_m\right]d^4x.
\end{equation}
The action above depends explicitly in $n+1$ quantities, namely the metric $g_{\mu\nu}$ and the auxiliary fields $\alpha_i$. The equations of motion for the fields $\alpha_i$ can be obtained through the variational method by taking a variation of the action with respect to these fields. The set of equations of motion for $\alpha_i$ can be conveniently expressed in terms of a matrix equation $M\textbf{x}=0$ of the form
\begin{equation}
    \begin{pmatrix}
        f_{\alpha_1\alpha_1} & \cdots & f_{\alpha_1\alpha_n} \\
        \vdots & \ddots & \vdots \\
        f_{\alpha_n\alpha_1} & \cdots & f_{\alpha_n\alpha_n}
    \end{pmatrix}
    \begin{pmatrix}
        X_1-\alpha_1 \\
        \vdots \\
        X_n-\alpha_n
    \end{pmatrix}
    =0,
\end{equation}
where the subscripts $\alpha_i$ denote derivatives of the function $f\left(\alpha_1,...,\alpha_n\right)$ with respect to $\alpha_i$. The matrix equation above features a unique solution if and only if the determinant of the matrix $M$ is non-zero, i.e., $\det M\neq 0$. In such a case, the unique solution for the matrix system above if $\textbf{x}=0$, i.e., $\alpha_i=X_i$ for all $i$. Replacing these solutions back into Eq. \eqref{eq:actionST}, one recovers Eq. \eqref{eq:action}, thus proving the equivalence between the two actions. Note that if $\det M=0$, the solution of the matrix system is not unique and thus the equivalence between the two representations is not guaranteed.

Introducing the definitions of the scalar fields $\phi_i$ and the scalar interaction potential $V\left(\phi_1,...,\phi_n\right)$ as
\begin{equation}\label{eq:deffields}
    \phi_i=\frac{\partial f}{\partial X_i},
\end{equation}
\begin{equation}\label{eq:defpotential}
V\left(\phi_1,...,\phi_n\right)=\sum_{i=1}^n \phi_i X_i-f\left(X_1,...,x_n\right),
\end{equation}
one can rewrite the action in Eq. \eqref{eq:actionST} in the more convenient form
\begin{equation}\label{eq:actionST2}
S=\frac{1}{2\kappa^2}\int_\Omega\sqrt{-g}\left[R+\sum_{i=1}^n \phi_i X_i-V\left(\phi_1,...,\phi_n\right)+2\kappa^2\mathcal L_m\right]d^4x.
\end{equation}
The action in Eq. \eqref{eq:actionST2} depends also on $n+1$ independent quantities, namely the metric $g_{\mu\nu}$ and the scalar fields $\phi_i$. The equations of motion for the scalar fields are obtained by taking a variation of the action with respect to $\phi_i$, from which one obtains
\begin{equation}
    \phi_i-V_{\phi_i}=0,
\end{equation}
where we defined $V_{\phi_i}\equiv \partial V/\partial \phi_i$. On the other hand, the field equations can be obtained by taking a variation of the action with respect to the metric, from which one obtains
\begin{equation}\label{eq:fieldST}
    G_{\mu\nu}-\frac{1}{2}g_{\mu\nu}\left(\sum_{i=1}^n \phi_i X_i-V\left(\phi_1,...,\phi_n\right)\right)
    +\sum_{i=1}^n H_{\mu\nu}^{(i)}|_{f_{X_i}=\phi_i}=8\pi T_{\mu\nu}.
\end{equation}
The field equations in Eq. \eqref{eq:fieldST} for the scalar-tensor representation are equivalent to the ones obtained previously in Eq. \eqref{eq:field} for the geometrical representation. This can be verified by introducing the definitions of the scalar fields $\phi_i$ and scalar potential $V$ from Eqs.\eqref{eq:deffields} and \eqref{eq:defpotential} into Eq. \eqref{eq:fieldST} to recover Eq. \eqref{eq:field}. Thus, the set of junction conditions for the scalar-tensor representation is equivalent to the one from the geometrical representation, albeit some of the conditions being written in terms of different quantities, namely the scalar fields $\phi_i$.

One of the main features of the scalar-tensor representation obtained through the methods outlined above is the fact that the scalar fields $\phi_i$, despite being independent quantities with their own equations of motion, are closely related to the original scalar quantities $X_i$. Indeed, a consequence of the requirement that $\det M\neq 0$, necessary for the scalar-tensor representation of the theory to be well-defined, is that the relationships between the scalar fields $\phi_i$ and the scalar quantities $X_i$ are invertible, i.e., one can write $\phi_i=\phi_i\left(X_1,...,X_n\right)$, as well as $X_i=X_i\left(\phi_1,...,\phi_n\right)$. The invertibility of the relationships between $\phi_i$ and $X_i$ implies that the junction conditions for the scalar fields $\phi_i$ can be extracted from the junction conditions for the scalars $X_i$. In order for the potential function $V\left(\phi_1,...,\phi_n\right)$ to remain regular in the distribution formalism, it is necessary that it does not feature any factors $\delta^2\left(l\right)$. Given that $\phi_i=\phi_i\left(X_1,...,X_n\right)$, the regularity of $V\left(\phi_1,...,\phi_n\right)$ is thus equivalent to the regularity of $f\left(X_1,...,X_n\right)$, thus giving rise to the same junction conditions as the immediate junction conditions obtained in Sec. \ref{sec:juncgen}. On the other hand, the presence of any differential terms in $H_{\mu\nu}^{(i)}$, which in the scalar-tensor representation correspond to differential terms in $\phi_i$, induce the same differential and coupling junction conditions in $\phi_i=\phi_i\left(X_1,...,X_n\right)$ as originally introduced in the geometrical representation, as well as the contributions to the stress-energy tensor of the thin-shell. 

The junction conditions for the scalar fields $\phi_i$ can also be obtained in a case-by-case basis. Indeed, for a given explicit form of the function $f\left(X_1,...,X_n\right)$ or, equivalently, a given explicit form for the potential $V\left(\phi_1,...,\phi_n\right)$, each scalar field $\phi_i$ depends on some subset of $X_i$ which varies depending on the form of the function considered. Thus, if the junction conditions require that the quantities $X_i$ are continuous, i.e., $\left[X_i\right]=0$, this implies immediately that $\left[\phi_i\right]=0$, or vice-versa. Note that in such particular cases, the requirement $\det M\neq 0$ must be verified independently to guarantee that the scalar-tensor representation is well-defined.

\section{Conclusions}\label{sec:concl}

In this work, we have considered a theory of gravity depending on an arbitrary number of scalar quantities $X_i$, i.e., the action of such a theory is described by a function $f\left(X_1,...,X_n\right)$, and we have analyzed the junction conditions of such a theory for several scalar quantities of interest. Our analysis includes the general methods to derive the junction conditions for an arbitrary action for both smooth matching and matching with a thin-shell, methods on how to simplify the general set of junction conditions by considering particular forms of the action, a description of the situations where double gravitational layers may arise, and an indication on how these junction conditions are modified if one introduces a dynamically equivalent scalar-tensor representation of the same theory.

Regarding the general sets of junction conditions, obtained when the function $f\left(X_1,...,X_n\right)$ is assumed to be analytic, i.e., that admits a Taylor series expansion up to any order with non-vanishing coefficients, we have characterized the junction conditions in three distinct groups: (i) immediate conditions, which arise from the imposition of regularity of the geometrical and matter quantities of interest of the theory and their respective derivatives; (ii) differential conditions, arising from imposing the regularity of terms proportional to higher-order partial derivatives of the function $f\left(X_1,...,X_n\right)$; and coupling conditions, which are induced when a non-minimal coupling between different scalars $X_i$ is present in the action. Our results indicate couplings between different scalars induce junction conditions similar to the differential junction conditions of each coupled scalar independently, but applied to the analogous quantity for the scalar it is coupled to. The same is true about the contributions to the stress-energy tensor of each scalar quantity and the ones induced by their coupling.

For particular forms of the function $f\left(X_1,...,X_n\right)$ for which some of the coefficients of its Taylor series vanish, we have show than the general set of junction conditions can be simplified. In particular, coupling junction conditions can be eliminated from the general set by considering an action for which no products between functions of different scalars are present. Furthermore, by decreasing the power dependency of the function $f\left(X_1,...,X_n\right)$ in some scalar $X_i$, the differential junction conditions can also be eliminated. Finally, if the function $f\left(X_1,...,X_n\right)$ is linear and decoupled for some scalar $X_i$, any immediate junction conditions arising from the arbitrary dependency of the action in that scalar can also be removed. In such a case, the only immediate junction conditions that remain in the set are the ones arising from the regularity of the geometrical and matter quantities themselves. Our results also indicate that, for the particular case that the function $f\left(X_1,...,X_n\right)$ is linear in every scalar $X_i$, the junction conditions reduce to forms qualitatively similar to those of the Israel or Darmois junction conditions, depending on the existence of restrictions on the stress-energy tensor of the thin-shell, like it happens e.g. for the scalars $T_{\mu\nu}T^{\mu\nu}$ and $R_{\mu\nu}T^{\mu\nu}$.

For certain particular forms of the function $f\left(X_1,...,X_n\right)$ for which differential junction conditions are removed from the set but second-order derivatives of the scalars $X_i$ are still present, one may find that double gravitational layers arise, i.e., singular contributions to the stress-energy tensor proportional to the distribution $\nabla_\mu\delta$. We have shown that such a situation is possible as long as the regularity scalar $X_i$ does not force the stress-energy tensor to be regular. We have also shown that the coupling between different scalars $X_i$ affects the contributions to the stress-energy tensor of the double gravitational layer in an analogous way as it affects the contributions to the stress-energy tensor of the thin-shell.

Finally, we have introduced the methods to construct a dynamically equivalent scalar-tensor representation of any theory $f\left(X_1,...,X_n\right)$ via the introduction of $n$ scalar fields $\phi_i$ and a scalar interaction potential $V\left(\phi_1,...,\phi_n\right)$. Such a scalar-tensor representation is only well defined if the relationship between the scalars $X_i$ and the scalar fields $\phi_i$ is invertible, which consequently implies that this representation is unique. For such a case, one verifies that the field equations in both representations are perfectly equivalent, and thus the junction conditions remain the same, albeit being written in terms of different quantities, i.e., they are given in terms of the scalar fiels $\phi_i$ if one considers the scalar-tensor representation, instead of the scalars $X_i$ if one considers the geometrical representation. The equivalence between these two sets of junction conditions can be assessed via the invertible relationship between these quantities.

The junction conditions are particularly important in the current scenario as they allow one to develop new alternative solutions for compact objects that could be tested in astrophysical scenarios by the current observations of gravitational wave signals, shadows, and astrometric measurements in the galactic center. The test of such models built using alternative theories of gravity as a foundation serve as indirect tests to these theories and could potentially shed light not only on the nature of ultra-compact objects but also on the fundamental understanding of gravity. We hope that this manuscript successfully suppresses the literary gap of junction conditions in modified theories of gravity and contributes to an advance in the study of compact objects in alternatives to GR.

\begin{acknowledgments}
We acknowledge Francisco S. N. Lobo for useful comments and suggestions. JLR acknowledges the European Regional Development Fund and the programme Mobilitas Pluss (MOBJD647), and project No.~2021/43/P/ST2/02141 co-funded by the Polish National Science Centre and the European Union Framework Programme for Research and Innovation
Horizon 2020 under the Marie Sklodowska-Curie grant agreement No. 94533. 
\end{acknowledgments}


\end{document}